\documentclass[aip, preprint, superscriptaddress, amsmath, amssymb]{revtex4-2}

\usepackage{color}
\usepackage{titlesec}
\usepackage{caption}
\usepackage[labelformat=simple]{subcaption}
\usepackage{graphicx}
\usepackage{amssymb}
\usepackage{amsmath}
\usepackage[hidelinks=true]{hyperref}
\hypersetup{
		colorlinks = true,
		urlcolor = blue,
		linkcolor = blue,
		citecolor = blue
	}

\begin{document}

\title{New Materials, New Functionalities: Molecular Beam Epitaxy of Ultra-High Conductivity Oxides}

\author{Gaurab Rimal}
\email{gaurab.rimal@wmich.edu}
\affiliation{Department of Physics, Western Michigan University, Kalamazoo, MI, USA}

\author{Tanzila Tasnim}
\author{Brian Opatosky}
\affiliation{Department of Physics \& Astronomy, University of Delaware, Newark, DE, USA}

\author{Ryan B. Comes}
\affiliation{Department of Materials Science and Engineering, University of Delaware, Newark, DE, USA}

\author{Debarghya Mallick}
\author{Simon Kim}
\author{Rob G. Moore}
\affiliation{Materials Science Division, Oak Ridge National Laboratory, Oak Ridge, TN, USA}

\author{Seongshik Oh}
\affiliation{Department of Physics \& Astronomy, Rutgers, The State University of New Jersey}
\affiliation{Center for Quantum Materials Synthesis, Rutgers, The State University of New Jersey}

\author{Matthew Brahlek}
\email{brahlekm@ornl.gov}
\affiliation{Materials Science Division, Oak Ridge National Laboratory, Oak Ridge, TN, USA}
\newcommand{\red}[1]{\textcolor{red}{#1}}
\newcommand{\blue}[1]{\textcolor{blue}{#1}}
\newcommand{\green}[1]{\textcolor{green}{#1}}

\begin{abstract}

Understanding fundamental properties of materials is necessary for all modern electronic technologies. Toward this end, the fabrication of new ultrapure thin film materials is critical to discover and understand novel properties that can allow further development of technology. Oxide materials are a vast material class abound with diverse properties, and, therefore, harnessing such phases is critical for realizing emerging technologies. Pushing forward, however, requires understanding basic properties of insulating, semiconducting and metallic oxides, as well as  the more complex phases that arise out of strong electronic correlations unique to this class of materials. In this review, we will focus on one of the unique aspects of oxides: the ultra-high conductivity metallic state, which can be a critical component for future all-oxide microelectronics such as low-loss interconnects and gate-metals, spintronics, as well as future quantum technologies that employ emergent magnetic or superconducting ground states. Like most oxides, a critical challenge to understanding and ultimately integrating high-conductivity metals into new technologies is the ability to synthesize high-quality materials. Therefore, we will frame the discussions in the context of epitaxial film growth via molecular beam epitaxy (MBE), which has provided insights into the electronic behavior of specific materials while providing samples with unprecedented quality. We will highlight and underscore how MBE has enabled developments and deeper understanding of their properties and how it plays a critical role in the future of this unique class of materials.

\end{abstract}

\maketitle

\section{Introduction}
 Ever changing improvements to the landscape of modern technology has been closely intertwined with fundamental understanding of the electronic, magnetic and optical properties as well as the ability to seamlessly integrate dissimilar materials and create structures on the nanoscale. This can be traced back to the early electronics industry where quantum mechanical models of the electronic properties of metals and insulators and the intermediate phases enabled the creation of the first prototype field-effect transistor, which utilized depletion or accumulation of the electronic charge to switch between conducting and non-conducting states. The ability to integrate materials like metals and dielectrics at heterointerfaces with these simple semiconductors enabled the creation of the first multi-transistor integrated circuit, which is the forefather of all major electronic devices in use today. In parallel with the development of the integrated circuit, researchers sought materials with more diverse bandgaps and band offsets to create and control optical responses. This led to major breakthroughs in the processing and adaptation of binary semiconducting materials such as the III-V (GaAs, InAs, AlN) and II-IV compounds (CdTe, ZnSe), which are more challenging to deposit as films. This is largely due to the fact that there are multiple elements which have dissimilar physical properties, such as disparate volatility of the constituent elements as well as the resulting compound, and the often toxic or pyrophoric nature of the elements. This early work in the 1950-70s resulted in a deeper understanding of material properties, growth and processing, as well as the development of ultrahigh vacuum technologies, and resulted in the incarnation of the modern molecular beam epitaxy (MBE) method \cite{Cho1975}.

Since the first integrated circuit decades ago, the technological landscape has dramatically changed. Centralized data processing units are surrounded by vast preferential components that cumulatively make  modern technologies integrate into our daily life. This ranges from diverse input-output technologies that enable interaction, such as displays, cameras, accelerometers and optical sensors, to volatile and non-volatile memory elements, and batteries that enable mobile computers that fit in our pockets and on our wrists. The heart of these are processor units that now contain billions of transistors. There has, moreover, recently been diversification in processing elements with the rise of GPU that are critical to the rapid development of generative artificial intelligence. However, the electronics industry, namely those that involve semiconductors, is facing a major challenge from the potential end of Moore's law. This law, which roughly states that the density of transistors doubles every 18 months with nominally fixed cost, has begun to saturate and will cease to exist when the size of transistors reach the atomic limit. Even now, electronic devices have components with feature size below 10 nm \cite{chen_sub-10_2021}, which is approaching the lowest values that can be achieved for current materials and device motifs. Although this seems somber, this is, in fact, an exciting technological epoch as industry expands away from a monoculture. To meet these current challenges researchers and engineers have begun to look at new and diverse motifs for computation that rely on aspects of the electron other than charge, such as spin or valley degrees of freedom in spintronics, phase change based novel switching mechanism such as neuromorphic computing, and fundamentally different paradigms such as quantum computing. A key commonality among all these is the need for \textit{new materials with new functionalities}. Additionally, new materials can have dramatic impact on the auxiliary components and can improve interfacing, robustness and lifetime, as well as energy efficiency.  

Typical electronic devices are comprised of semiconductors, insulators and metals. The semiconductor is at the heart of most devices since it can be readily switched between an insulating state where no electrical current can flow and a metallic state where current is free to flow. For the field-effect device specifically, this conductive and non-conductive state corresponds to binary 1s and 0s. The modulation of the semiconductor is mediated by atomic scale interfaces with metals and insulators. The switching occurs by applying an electric field in a capacitor-like geometry where a gate-metal is isolated from semiconducting channel by an insulator. In current generations these are composed of semiconductors like Si, GaAs and GaN, which can be chemically doped to become metals and serve as back gates and the separating insulator layer is most simply SiO$_2$. The insulator can vary based on the length of the device, for example high-k dielectrics such as hafnium oxide made their debut in the mid 2000s \cite{robertson_high-k_2015}. Simple metals like Cu are integrated as metallic contacts such as interconnects \cite{rosenberg_copper_2000, grill_progress_2014,kim_addressing_2024} which convey electrical signals to and from these components. Other passive components such as resistors and capacitors are similarly made of different materials. Similarly, specialized metals make their appearance in a wide array of other devices. For example, all flat-panel devices are cladded by a transparent conductive oxide (typically indium tin oxide, abbreviated as ITO) that enables electrical current to pass through which modulates the light emission while remaining transparent to visible light. Although magnetic hard drives are currently being replaced by solid state drives that rely on floating-gate transistors, magnetic metals are critical components for storing information in which the direction of the magnetization (up or down) corresponds to 0 or 1 bit state. 

Over the past few decades there has been a push for all-oxide electronics \cite{Ramesh2008} where oxides can be assembled as the semiconductor, the insulators, as well as the metal. A key specialty about oxides is their vast chemical and structural diversity as well as their electronic and magnetic properties: there are examples of oxides which are good metals, semiconductors, insulators, and typically exhibit a plethora of different auxiliary behaviors. These diverse behaviors offer many functionalities beyond simple semiconductors and can be used to address many technological problems. One of the central novelties of oxides is that the basic functional element, the electron, essentially becomes more complex and rich in properties. For example, charge current in a metal is carried by the valence electrons. However, all the electrons interact electrostatically through the Coulomb interaction with each other as well as the nuclei. This creates an unsolvably complex problem. Through a blessing of nature, however, in most materials this highly complex many-body state of the electrons can be replaced by quasi-electrons that are essentially non-interacting. The only difference is these quasi-electrons have slightly different masses, but, otherwise, behave like classical objects. In oxides, in contrast, this is not generally true because the cation metal atoms are bonded through oxygen ligands. When cations are transition metals the bonding is very localized, which, when coupled to the highly electronegative oxygen creates very weak band-overlap, thus making it hard for electrons to become delocalized. This gives rise to situations where novel magnetic phases as well as other strongly correlated phases such as high temperature superconductivity arise. Moreover, this often results in materials that should be metals and are in fact empirically found to be insulators due to it being energetically unfavorable to delocalize in the process of band formation, which is known as a Mott insulator \cite{Mott1968a}. Interestingly, there is often very subtle coupling that can drive oxides across such transitions, for example structural (Eg: vanadium oxides \cite{wu_design_2013,hu_vanadium_2023}), magnetic (Eg: colossal magnetoresistance in the manganites \cite{tokura_critical_2006}) or quantum critical transitions and high-temperature superconducting states (Eg: cuprates and nickelates \cite{keimer_quantum_2015,wang_experimental_2024}). 

Interestingly, in this view of oxide physics, a high conductivity metallic phase is in fact quite strange. Firstly, having a metallic state in general is quite strange in these types of materials since the band overlap is quite weak and correlations are naturally quite strong. However, chemistry can overcome this weak band overlap. Specifically, the orbitals for the 4d and 5d transition metals as well as p-block elements are much more delocalized. This results in larger band overlap and increased screening, which effectively weakens Coulomb repulsion to the point where dispersive bands can form. Secondly, the level of band overlap is critical for determining the quasi-electron effective mass. Semiconductors and simple metals typically have masses that are similar or smaller than that of the bare electron effective mass. In heavy semiconductors such as InSb there are very high orbital overlaps that result in effective masses that are orders of magnitude smaller than electrons. The low effective mass is critical for high conductivity metals since the conductivity scales inversely with the effective mass. Thus, oxides, in the rare cases that they are metals, should have intrinsically low conductivities. 

\begin{table}[ht]
\centering
\caption{Materials with their room temperature resistivity, main properties and potential applications. Resistivity values for MBE-grown materials are for the thickest films and are considered at the bulk limit. }
\begin{tabular}{|p{12ex}|p{12ex}|p{10ex}|p{10ex}|p{26ex}|p{25ex}|}
    \hline
     Type & Material & \multicolumn{2}{c|}{Resistivity ($\mu \Omega cm$)} & Properties & Applications \\
     & & Bulk SC & MBE &  & \\
     \hline
     Rocksalt & NbO & 21 \cite{hulm_superconductivity_1972} & -- & Superconductor & Gates \\
     
     \hline
     
     & CrO$_2$ & 60 \cite{heffernan_role_2016} &  -- & {Half metal} & Spintronics   \\
     
      & IrO$_2$ & 30 \cite{butler_crystal_1971} & 26 \cite{kawasaki_engineering_2018} & {Large SOC, Nodal line semimetal} & Spintronics   \\
     
     Rutile & RuO$_2$ & 28 \cite{butler_crystal_1971} & 56 \cite{nunn_solid-source_2021} &  {Superconductor, Altermagnet} & Spintronics \\

     & OsO$_2$ & 15 \cite{yen_growth_2004} & -- &  Nodal line semimetal &  Spintronics  \\
     
     \hline
     
     & ReO$_3$ & 9.2 \cite{quirk_singular_2021} & -- &  Flat bands, high mobility & Spintronics \\
     
     & SrVO$_3$ & 26 \cite{chamberland_alkaline-earth_1971} & 30 \cite{moyer_highly_2013} & Translucent, strongly correlated & Transparent conductor \\
     
     Perovksite & SrMoO$_3$ & 5 \cite{nagai_highest_2005} & 24 \cite{takatsu_molecular_2020} & Translucent & Transparent conductor \\
     
     & SrNbO$_3$ & 330 \cite{xu_red_2012} & 80 \cite {palakkal_off-stoichiometry_2025} & Translucent, topological  & Transparent conductor  \\

     & LaNiO$_3$ & 90 \cite{guo_antiferromagnetic_2018} & 78 \cite{wrobel_comparative_2017} & Strongly correlated  & Interfaces,  precursor to superconductor \\
     
     \hline
     
     & PtCoO$_2$ & 2.1 \cite{Mackenzie2017} & 4.5 \cite{song_surface_2024}  &  {Electron viscosity, surface magnetism, anisotropy} & Interconnect, spintronics \\
     
     Delafossite & PdCoO$_2$ & 2.6 \cite{Mackenzie2017} & 3.6\cite{Brahlek2019} &   {Electron viscosity, surface magnetism, anisotropy} & Interconnect, junction, gates \\
     
     & PdCrO$_2$ & 5 \cite{Mackenzie2017} & 60\cite{song_surface_2024} &  {Magnetic, correlated, anisotropy} & Interconnect, gate  \\
     
     \hline
\end{tabular}
\label{table:material_properties}
\end{table}

As shown in Table \ref{table:material_properties} the binary oxides demonstrate these general characteristics of band-filling and correlations in regards to the formation of the metallic phase. For the 3D transition metal oxides, there are only a few high-conductivity metals \cite{goodenough_metallic_1971}. The elements near the beginning and end of the 3d-block of the periodic table (Eg: Sc, Ti, and Zn), form insulators simply due to completely filled/unfilled bands. Materials near the center of the 3d-block such as NiO are classical examples of Mott insulators due to small band width, which is due to the highly localized orbitals. Moreover, for the non-insulating examples, the low conductivity demonstrates the high effective masses that also derive from the narrow bandwidth. Moving to the 4d transition metal oxides, there are more true metallic states such as RuO$_2$ and some of the niobium oxides. This demonstrates the effect that the larger 4d orbitals has on the bonds to the oxygen and the corresponding larger bandwidth and weaker effects of correlations. This trend is even more dramatic in the 5d transition metal oxides where materials like rhenium oxide, osmium oxide and iridium oxides are high conductivity metals. 

\begin{figure}[ht]
    \centering
    \includegraphics[width=0.8\textwidth]{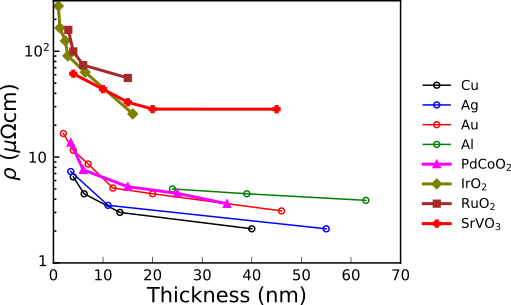}
    \caption{Comparison of room temperature resistivity of MBE-grown high conductivity oxides with conventional metal films. Adapted from Rimal et al \cite{Rimal2021}. }
    \label{fig:resistivity-thickness}
\end{figure}

As such, understanding the basic physics of oxide metals is of critical importance both from the perspective of basic condensed matter physics but also for technological applications. Central to this is the growth of clean thin film materials for which MBE is a critical tool that enables spanning basic and applied research. Oxide MBE is generally more challenging compared to traditional semiconductors. However, steady progress over the past several decades has now allowed realization of high quality epitaxial films of high-conductivity metallic oxides. We will consider materials with bulk room temperature resistivity below about 100 $\mu \Omega cm$  to define ultra-high conductivity. This low resistivity as well as the scaling of resistivity with thickness in the thin-film limit is comparable to the best metals, as shown in Figure \ref{fig:resistivity-thickness}. This places these oxides in the same regime as simple metals, thereby providing large potential as components in electronics and emerging technologies. Moreover, many of these materials exhibit novel emergent phenomena which make it even more relevant for nascent technologies. 

In this review, we will focus on the key materials properties and the developments of MBE that have brought the field to the current positions, as well as key challenges to push the field forward in the coming decade. To accomplish this we first give a brief introduction to the MBE growth and its role in uncovering intrinsic properties of oxides materials over the past four decades. We then discuss individual oxide materials that exhibit high-conductivities at room temperature. We will focus our discussion on the following material systems: the binary rocksalts (VO, NbO), rutiles (CrO$_2$, RuO$_2$, IrO$_2$ and OsO$_2$), $ABO_3$ perovksites (Sr/CaVO$_3$, SrMoO$_3$, SrNbO$_3$, LaNiO$_3$ and tungsten bronze phases) and finally the ABO$_2$ delafossites (PdCoO$_2$, PdCrO$_2$, and PtCoO$_2$). Each section will be dedicated to a particular structural class of ultra-high conductivity oxide. The individual sections will focus on particular chemistry within a structural class and will be organized by giving basic physical properties of each material, review work on the growth of that particular material with an emphasis on work done by MBE, discuss novel functional properties, and close by presenting questions regarding how MBE can be used to push the quality of samples and interesting open questions. We then conclude with a detailed discussion of general open questions and opportunities in the field of ultra-high conductivity oxides.

\section{Oxide molecular beam epitaxy for ultra-high conductivity metals}
MBE has become the premier epitaxial growth technique for thin film materials, in that it provides the highest level of control for epitaxial growth with atomic scale accuracy. The development of MBE started in the late 1950s through to the 1980s as a method to grow semiconductor materials as well as metals \cite{Cho1975,schuller_magnetic_1999}. Conceptually, MBE is very simple in that it relies on thermal evaporation and sublimation of highly purified elemental sources which then adsorb onto a single crystal substrate which acts as a template from which the film subsequently grows. For thermal evaporation to work, MBE is performed in vacuum at extremely low pressures (called ultra-high vacuum (UHV)), which are typically less than $10^{-7}$ Torr (compared to 760 Torr of atmospheric pressure). UHV is critical as it enables long mean free paths of the elements once evaporated, thus forming the "molecular beams". This reduces source-to-source contamination and focuses all chemical interactions/reactions to happen on the growing surface rather than in the vapor phase. The UHV environment is also critical as it nearly eliminates any contamination of the growing surface with the background gases in the chamber, thus further reducing extrinsic contamination. Finally, one of the critical aspects unique to MBE is the ability to incorporate in-situ and real time probes of the growing surface. Over the long history of MBE, this has included a list of techniques such as electron and x-ray diffraction, optical and x-ray spectroscopies, and others. The primary technique that is ubiquitous to MBE is reflection high-energy electron diffraction (RHEED). RHEED utilizes a high energy electron-beam in the order of 10 keV in a glancing geometry. The resulting diffraction pattern is exquisitely sensitive to the crystalline properties and morphology of the top-most layer of atoms. This realtime feedback is critical to being able to rapidly identify optimized growth conditions for new material systems. These aspects combine to make an extremely flexible technique that is applicable across myriads of material systems and is the premier technique for creating materials of the highest quality. Despite this long history, there are still advances to the MBE technique which are being driven by the exploration of new materials, pushing the limits of growth to achieve even higher quality.

\begin{figure}
    \centering
    \includegraphics[width=0.9\textwidth]{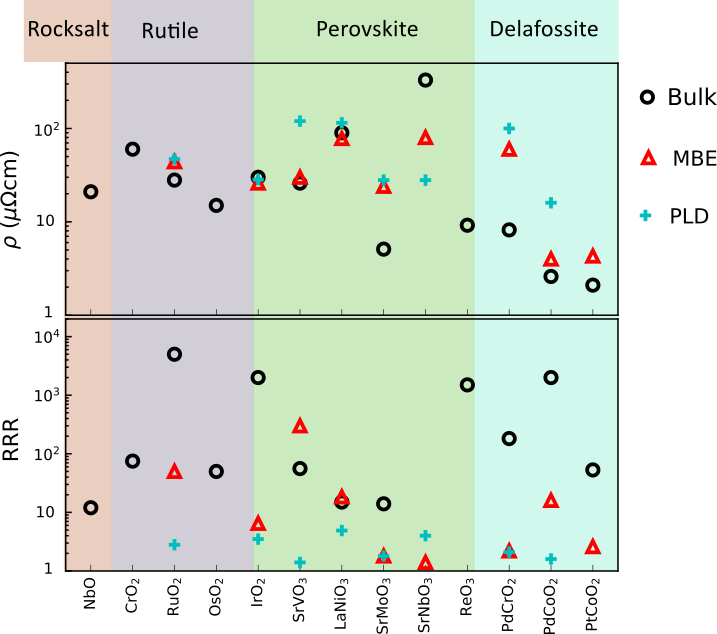}
    \caption{Comparison of resistivity and residual resistivity ratio for high conductivity oxides for bulk, MBE and PLD grown epitaxial thin films. }
    \label{fig:bulk-thickness_fom}
\end{figure}

One of the major advantages of utilizing MBE growth for ultra-high conductivity oxides is its ability to control defects that helps provide insights into the fundamental properties. This is extremely challenging since most oxides are not simple line compounds, where the compound has exact stoichiometry, and thus many oxides are difficult to grow with perfect stoichiometry during synthesis. MBE in particular has additional challenges. The prototypical material for MBE growth is GaAs, which as a line compound will tend to only form GaAs during growth. For MBE growth this enables an extremely simple strategy called \textit{adsorption controlled growth} based on the existence of a growth window. Adsorption controlled growth is possible since excess As can be supplied for a wide range or window in temperature and pressure. This is due to the large difference in vapor pressures among As, Ga, and the desired GaAs: when an over-pressure of As is supplied to the growing surface, excess As is not incorporated into the crystal and desorbs off the surface, which leaves extremely pure GaAs with near perfect stoichiometry. 

Oxides in general are more challenging compared to elemental and binary semiconductors. This is because adsorption controlled growth is not applicable for most oxide materials due to their typically high sublimation temperatures as well as other challenges.  Some of the key challenges are as follows: there are often many phases that are in close proximity in the phase diagram and these phases are typically comprised of multiple oxidation states, especially for transition metals, and achieving a targeted oxidation state or crystalline phase can be difficult. This is made more difficult by the fact the MBE growth occurs in the reducing UHV conditions. However, since the mid 1980s scientists and engineers have come up with novel strategies and methods to grow high quality oxides by MBE. Introducing background gas into the reactor at pressures up to 10$^{-5}$ Torr during growth gives control over oxidation for many elements. Development of reactive plasma sources, distilled ozone and supplying prereacted species as metal organics have enable greater control. Interestingly, for many oxides this highly oxidizing condition has created a secondary problem for specific elements where the oxide is in fact more volatile than the individual elements. This occurs for example in the cases of Ir, Ru, Sn, but depending on the specific oxidation state can occur for other elements. The increased volatility of the oxide causes reevaporation of the element off the growing surface. Although this may seem to be critical challenge, it has been repurposed to solve an additional challenge of controlling stoichiometry and can be applied as a variant of adsorption controlled growth. 

The case for ternary and quaternary oxides is even more complicated, as there are often no broad growth windows, thus the stoichiometry of the elements require manual control. Specifically, for the case of the $ABO_3$ perovskites, the ratio of $A$ and $B$ need to be set manually. For MBE growths, this is a challenge since the accuracy of in-situ calibrations are only at the percent level, thus leaving films with potentially high levels of defects. Moreover, the strong oxidizing conditions are well-known to cause source oxidation, which subsequently causes the flux of the sources to drift during the growth, and thus the stoichiometry of the film also changes across the thickness. To address these challenges MBE growers have come up with a myriad of solutions. This includes detailed protocols for accurate source calibrations utilizing in-situ diagnostics such as quartz crystal microbalance (QCM) used in conjunction with RHEED. 

Other routes to create growth windows have been developed. As mentioned above, this includes controlling oxidation of one of the metals (for example Pb in PbTiO$_3$ or Ru in SrRuO$_3$) to create an oxide of different volatility. This enables oversupplying one of the elements where the excess resublimes from the surface before it is incorporated in the film, thus leaving near perfect stoichiometry. More recently, incorporating metal organics have resulted in routes to stoichiometric oxide films in what has become known has hybrid or metalorganic MBE \cite{choudhary_atomically_2025}. Here, the cation being bound to organic ligands makes it naturally volatile which enables any excess flux supplied to the growing surface to evaporate back into the gas phase. More importantly, for certain cases the breakdown of this molecule into the metal or metal oxides is often tied to the supply of the other cation. This opens a growth window and results in near perfect stoichiometry. Moreover, the utilization of metal-organics enables incorporating refractory and near refractory elements to be used in MBE, thereby circumventing high temperature MBE cells or electron-beam evaporators. Altogether, these innovations have enabled creating thin films with low defects which allows better control over the separation of intrinsic and extrinsic properties. As a result, epitaxial films grown using MBE compare to the best single crystals as demonstrated by comparable resistivity and residual resistivity ratio (RRR) values as demonstrated in Figure \ref{fig:bulk-thickness_fom}. For more information regarding MBE growth of oxide materials, we would like to point the reader to the following review articles \cite{rimal_advances_2024,Brahlek2018,Brahlek2020,choudhary_atomically_2025,Chambers2000a}.

In the following, we will focus on specific examples of ultra-high conductivity oxides, and how their fundamental properties are impacted and can be better understood through epitaxial films. 

\begin{figure}
    \centering
    \includegraphics[width=0.5\textwidth]{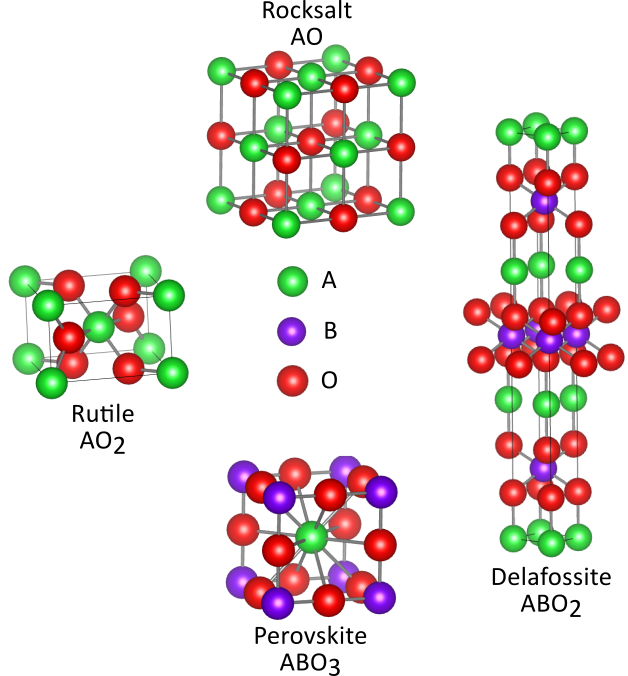}
    \caption{ Crystal structures for the common ultra-high conductivity oxides. }
    \label{fig:structures}
\end{figure}

\subsection{High conductivity binary rutiles }
Rutiles are a class of materials that get their name from the stable  form of the material TiO$_2$. The rutile system has a tetragonal crystal structure with symmetry of $P4_2/mnm$. The metal cation is in 4+ valence state and octahedrally coordinated by oxygen: the octahedra are edge-shared along the c-axis and corner connected along a-b axis as shown in Figure \ref{fig:structures}. This connectivity manifests in crystals growing fast along the c-axis, which results in needle-like growth in bulk \cite{rogers_crystal_1969}, and gives rise to highly anisotropic optical \cite{sinton_birefringence_1961} and electronic properties \cite{hollander_anisotropic_1960}. The optical response of TiO$_2$ has exciting potential in optoelectronic applications \cite{ge_review_2023}. Many additional transition metals oxides crystallize in the rutile structure: some well known rutiles include the ultra-wide bandgap material GeO$_2$ \cite{robertson_electronic_1979}, the semiconductors TiO$_2$ and SnO$_2$ (In and F doped SnO$_2$ are used as transparent conductors in electronic devices) \cite{ellmer_past_2012},  VO$_2$ which undergoes a first order metal-to-insulator transition \cite{hu_vanadium_2023} and the half metal CrO$_2$ \cite{Coey2002}. 

Of interest here are the metallic rutiles, which include RuO$_2$, IrO$_2$, CrO$_2$ and OsO$_2$, which all exhibit room temperature resistivities below 100 $\mu\Omega$cm. Given the metallic ground state,  strongly correlated nature, and strong spin-orbit coupling, the metallic rutiles exhibit  emergent behavior such as superconductivity in strained RuO$_2$ \cite{uchida_superconductivity_2020} and mysterious magnetism in RuO$_2$ \cite{smejkal_crystal_2020}. The large spin-orbit coupling of IrO$_2$ has been predicted and shown to be a nodal line semimetal with large intrinsic spin Hall effect which has great potential in spintronics \cite{sun_dirac_2017}. These make the metallic rutiles of interest for fundamental studies and possible applications. The simple binary nature, moreover, makes the rutiles attractive as a testbed for investigations into the fundamental nature of materials. However, the growth of rutiles via MBE is still in its infancy, and is most severely limited by the choice of substrates available. Currently, TiO$_2$ is the only substrate available, and with lattice constants of a = 4.59 $\textrm{\AA}$ and c = 2.96 $\textrm{\AA}$, epitaxial mismatch is a major concern for growing rutile materials. There are also challenges associated with the existence of multiple oxidation states and associated issues with proper growth windows. Still, new insights and interesting discoveries have been unraveled through epitaxial engineering of some of these materials. We will discuss these by focusing on individual materials below. 

\subsubsection{IrO$_2$ } 
Rutile IrO$_2$ exhibits room temperature resistivity around 30 $\mu\Omega cm$ \cite{butler_crystal_1971}. Ir is in the expected 4+ valence state, leaving 5 unpaired electrons in the d-shell. In the octahedral crystal field the d-orbitals are split into the lower lying 3-fold degenerate $t_{2g}$ and higher 2-fold degenerate $e_g$ levels. This should result in a partially filled $t_{2g}$ manifold, and, therefore, IrO$_2$ is expected to be metallic. Correlation effects are expected to be small due to the spatially extended nature of the 5d orbitals, and SOC is strong for heavy elements such as Ir, which drives additional splittings of the subbands and changes the effective band filling \cite{zhang_recent_2024}. Interestingly, SOC combined with the nonsymmorphic symmetry in rutile IrO$_2$ was predicted to exhibit opologically protected Dirac nodal lines \cite{kahk_understanding_2014,kim_spin-orbit_2016,zhang_recent_2024}. These aspects create electronic states at the Fermi level that are strongly coupled to the spin states leading to strong spin Hall effect that can be used for spin-induced switching \cite{fujiwara_5d_2013,patton_symmetry_2023}. As such, the major driver of recent research efforts in IrO$_2$ has been its significant potential in spintronics: ab initio study by Sun et al demonstrated direct correspondence between Dirac Nodal Lines in IrO$_2$ and a large spin Hall conductivity of $-2.5 \times 10^4 \, \hbar/e$ \cite{sun_dirac_2017}. 

In studying the rutiles, bulk single crystals of IrO$_2$ were grown many decades ago: Ryden and Lawson conducted transport studies on IrO$_2$ bulk single crystals grown via chemical vapor transport, where they observed different growth variations such as needles oriented along [001] and [011] and plates along (011) and (100) depending on the temperature and oxygen flow rates \cite{Ryden1970}. Butler and Gilson also found frequent occurrence of twinning and layered surface defects in their CVT-grown rod-like IrO$_2$ single crystals \cite{butler_crystal_1971}. They indicated that resistivity at low temperature could be sensitive to growth conditions, and optimal conditions led to room temperature resistivity as low as 30 $\mu\Omega cm$ and RRR of around 1800 which demonstrates that bulk crystals can be reliably grown and are stable at ambient conditions. Further research on this materials was mainly limited as a testbed to understand the optical and electronic properties \cite{goel_optical_1981,lin_low_2004,de_almeida_electronic_2006}, as well as catalytic behavior for oxygen or hydrogen evolution reactions \cite{over_fundamental_2021}. 

The prediction of topological behavior in IrO$_2$ brought new enthusiasm into this material, and epitaxial films were deemed necessary to unsderstand the intrinsic behavior. As a result, high quality films of IrO$_2$ have been grown using sputtering \cite{patton_symmetry_2023}, PLD \cite{kesler_epitaxial_2024} and MBE \cite{kawasaki_engineering_2018,kawasaki_rutile_2018,nelson_dirac_2019}. There are several key challenges in growing IrO$_2$ films by MBE. The first is the low vapor pressure of Ir, which in a traditional MBE process requires an electron-beam evaporator: although the proper temperatures for evaporation can be reached by local electron heating, the source flux can vary widely over time and create density gradient in the film. The second challenge is the difficulty in oxidizing metallic Ir, which necessitates highly reactive sources such as ozone or atomic oxygen using an RF source. The third challenge occurs when Ir is oxidized: the oxide exhibits an associated increase in volatility compared to that of metallic Ir, which can cause loss of Ir from the surface. This challenge manifests in the PLD growth of various iridate thin films, where Ir metal clusters form in low O$_2$ pressure, and significant loss of Ir occurs at high O$_2$ pressure due to the formation of volatile IrO$_3$ \cite{kesler_epitaxial_2024}.  Finally, the fourth challenge arises due to lack of proper substrates: TiO$_2$ is the only compatible substrate that is readily available, but the lattice mismatch between IrO$_2$ (a = 4.498 $\textrm{\AA}$, c = 3.16 $\textrm{\AA}$) and TiO$_2$ (a = 4.59 $\textrm{\AA}$, c = 2.96 $\textrm{\AA}$) is moderate but still significant that can readily lead to defects and disorder. As a result, columnar grains with relatively high grain boundary resistance occur by the nucleation of misfit dislocation boundaries \cite{hou_microstructure_2017}. Grain formation and strain relaxation occur during the growth of the first few unit cells, with some of the misfit dislocations leading to extended defects. With all these challenges, the room-temperature resistivities for epitaxial films are close to the expected bulk single-crystal values, yet the low-temperature resistivity values do not follow bulk behavior, highlighting the existence of extrinsic disorder. Despite these problems, atomically flat and well ordered surfaces can be grown in a MBE process, and recent technological developments make the continued work on IrO$_2$ exciting.

MBE-grown IrO$_2$ films exhibit room temperature resistivity of 26 $\mu\Omega cm$ and have RRR of about 6 when the films are grown using Ir supplied by an electron-beam evaporator \cite{uchida_field-direction_2015,kawasaki_engineering_2018}. The low RRR implies that there are residual defects, which likely stem from non-ideal Ir:O stoichiometry, and defects imparted by the substrate. Highly crystalline and atomically smooth films have been grown that have low residual resistivity values and metallic behavior at low temperatures down to the ultrathin limit of 3ML (0.96nm), while thicker films exhibit residual resistivity around 4 $\mu\Omega cm$, the lowest of any IrO$_2$ films \cite{kawasaki_engineering_2018}. Additionally, these films  provided an ideal platform to measure the band structure using angle-resolved photoemission (ARPES) and study the dimensionality-induced electronic properties: it was found that in atomically thin IrO$_2$ films, the effective mass of quantum well sub-bands were enhanced as much as six times compared to the bulk \cite{kawasaki_engineering_2018} and these effects are responsible for the ultra-thin films being metallic \cite{kawasaki_rutile_2018}. This was made possible through the ability of MBE to control materials layer-by-layer for heterostructure engineering. MBE-grown films that were transferred in-situ for ARPES measurements also helped to experimentally demonstrate the Dirac nodal line \cite{nelson_dirac_2019}. The crystal orientation and directional dependence of Hall effects \cite{uchida_field-direction_2015,yang_coexistence_2025} were also enabled by MBE-grown films. 

Recent work has demonstrated that the metal-organic precursor Ir(acac)$_3$ can enable the growth of high-quality iridium oxide thin films by overcoming the limitations associated with elemental iridium sources \cite{nair_engineering_2023,nair_high-mobility_2024,choudhary_semi-metallic_2022,rimal_strain-dependent_2024}. Ir(acac)$_3$ is solid and can be directly sublimated using a effusion cell operating around 100 - 200 $^{\circ}$C \cite{nair_engineering_2023}. This simple hybrid MBE method offers enhanced chemical reactivity and a more stable flux of iridium compared to electron-beam evaporators, and allows for epitaxial film growth without the need for highly oxidizing conditions that often lead to reevaporation of volatile Ir oxides from the surface. This approach showed that epitaxial strain plays a pivotal role in stabilizing IrO$_2$ over metallic Ir during growth, offering a new strategy to control oxidation thermodynamics in systems with hard to oxide metals. 

 The large SOC in this material is a central point of interest in the spintronics community. The spin Hall effect was experimentally measured in IrO$_2$ by Fujiwara et al \cite{fujiwara_5d_2013} using a lateral spin-valve device consisting of a sputter-grown IrO$_2$ wire and Py/Ag/Py layered structure on a SiO$_2$/Si substrate, showing a noteworthy inverse spin Hall effect and its efficacy as a spin current detector. The spin Hall angle and resistivity reported are 0.04 and 8 $\mu\Omega cm$ for polycrystalline IrO$_2$, and 0.065 and 37.5 $\mu\Omega cm$ for amorphous IrO$_2$ at 300 K. Recently, spin Hall angle of 0.45 was found in epitaxial (001) IrO$_2$ films, which rises to 0.65 at 30 K \cite{bose_effects_2020}. Other experiments involving crystal orientation dependence and spin-orbit torque manipulations have also been carried out \cite{patton_symmetry_2023,yang_coexistence_2025}. The superb spin Hall angles of IrO$_2$ \cite{Sinova2015} show that this material is likely to be an important component of future electronics and spintronics and MBE-grown epitaxial films can yield superb quality that can be applied for future benchmarks.

\subsubsection{RuO$_2$}

Like IrO$_2$, RuO$_2$ was first studied for understanding rutile materials. Bulk single crystals of RuO$_2$ have room temperature resistivity of 28 $\mu \Omega cm$ with RRR about 5000 \cite{Ryden1970, butler_crystal_1971}. Ru in the 4+ oxidation state has 4 unpaired electrons in the d-shell, and, like IrO$_2$, the $t_{2g}$ bands are partially filled, making RuO$_2$ metallic. Most research on this oxide has now focused on several novelties that have been observed or predicted. Similar to IrO$_2$ films there are possibilities of topological states emerging in the band structure of RuO$_2$ \cite{jovic_dirac_2018}. Moreover, recently observed superconductivity in strained films has received quite a bit of attention since superconductivity is absent in bulk crystals. Another interesting aspect of RuO$_2$ includes the mysterious itinerant antiferromagnetism that has been reported in some but not all samples. 

\begin{figure}
    \centering
    \includegraphics[width=0.9\textwidth]{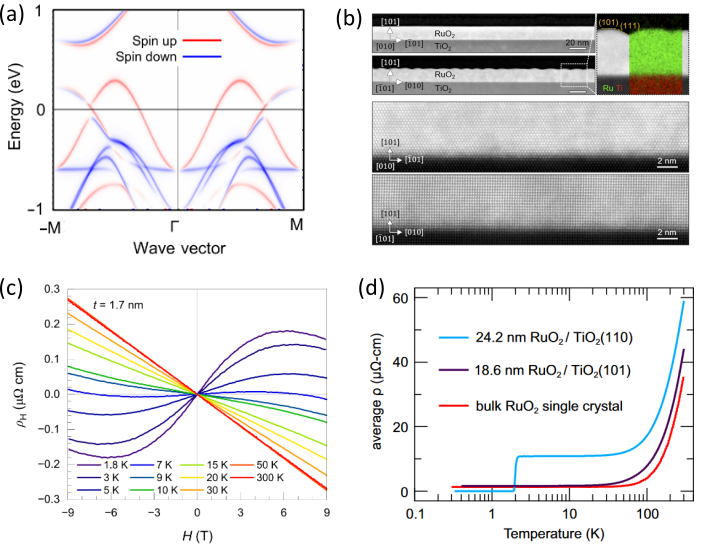}
    \caption{(a)  DFT calculated band structures for RuO$_2$ showing altermagnetic spin splitting. Taken from Fedchenko et al \cite{fedchenko_observation_2024}. (b) Scanning transmission electron microscopy for RuO$_2$ films on TiO$_2$ substrates grown using hybrid MBE. Taken from Nunn et al \cite{nunn_solid-source_2021}. (c) Anomalous Hall effect for thin films of RuO$_2$ grown using hybrid MBE. Taken from Jeong et al \cite{jeong_metallicity_2025}. (d) Superconductivity in strained RuO$_2$, taken from Ruf et al \cite{ruf_strain-stabilized_2021}. }
    \label{fig:RuO2}
\end{figure}

If antiferromagnetism can be stabilized in RuO$_2$ this can make it an interesting \textit{altermagnetic} candidate \cite{smejkal_crystal_2020}. Altermagnets are collinear antiferromagnets that intrinsically break time-reversal symmetry. This breaking of time-reversal symmetry stems from the fact that the crystalline unit cell contains two magnetic cations. Time reversal symmetry flips the relative spin direction of the spins on the two cations by $\mathrm{180^{\circ}}$. The spins can be translated arbitrarily by a combination of vectors from the basis of the crystalline unit cell. For most antiferromagnets this recovers the original unit cell, thus obeying time-reversal symmetry. For altermagnets, however, this is not the case and they explicitly break time-reversal symmetry, making altermagnetic candidates in some way similar to ferromagnets, yet without a net moment. The most exciting similarity to ferromagnets is the proposed large band splitting (see Figure \ref{fig:RuO2} and Smejkal et al \cite{smejkal_emerging_2022}). The key difference is that the band splitting is not homogeneous in momentum space, but rather \textit{alternates} across the Fermi surface. The rutile structure of RuO$_2$ fits the necessary criteria for such spin across the Fermi surface, and introducing antiferromagnetism should create a non-zero spin splitting. Although RuO$_2$ was long considered to be a Pauli paramagnet, neutron and x-ray scattering experiments found that this material under certain situations is a collinear antiferromagnet with an average magnetic moment of 0.5 $\mu_B$/Ru \cite{berlijn_itinerant_2017,zhu_anomalous_2019}. However, it is becoming evident that the ground state is in-fact non-magnetic. For example, muon spin relaxation ($\mu$SR), an extremely sensitive local probe of magnetism, failed to show ordering in bulk single crystals as well as epitaxial films \cite{hiraishi_nonmagnetic_2024,kesler_absence_2024}. However, experiments on epitaxial films showed that RuO$_2$ exhibits anomalous Hall effect \cite{feng_anomalous_2022,tschirner_saturation_2023} as shown in Figure \ref{fig:RuO2}, a sign that time reversal symmetry is broken \cite{Nagaosa2010}, yet is also possible that the anomalous Hall effect arises from small levels of magnetic defects, as was found in other altermagnetic candidates\cite{chilcote_stoichiometry-induced_2024}. ARPES and circular dichroism showed spin splitting in RuO$_2$ (110) films that could be ascribed to band-splitting\cite{fedchenko_observation_2024}, however spin-resolved ARPES \cite{liu_absence_2024} and optical conductivity \cite{wenzel_fermi-liquid_2025} studies failed to observe such splittings. The consensus in the community is that RuO$_2$ is intrinsically non-magnetic, and, therefore, not altermagnetic. However, under certain situations it may become weakly magnetic, which may be related to defects.

The ability to strain epitaxial thin films of RuO$_2$ revealed that MBE-grown samples on TiO$_2$ exhibit superconductivity, which is absent in bulk samples \cite{uchida_superconductivity_2020,ruf_strain-stabilized_2021,liu_absence_2024}. More specifically, it was found that MBE-grown RuO$_2$ films on (110)-oriented TiO$_2$ substrates exhibit a maximum superconducting transition temperature of 2 K (see Figure \ref{fig:RuO2}). However, for films grown on the other orientations of TiO$_2$ superconductivity was inhibited, and films grown on MgF$_2$ (110, 001) substrates were also not superconducting. The emergence of the superconducting state was found to be highly sensitive to the strain state of the film and the thickness. Uchida et al. performed detailed structural analysis which showed that epitaxial strain, particularly along the c-axis, leads to a shortening of Ru–O bonds and the softening of low-energy phonon modes, which they identify as essential to stabilizing superconductivity \cite{uchida_superconductivity_2020}. On the other hand, Ruf et al. combined MBE growth with in-situ ARPES and DFT calculations to reveal that anisotropic strain drives a redistribution of Ru 4d orbital occupancies, which enhances the electronic density of states and stabilizes superconductivity. They were also able to preclude interface effects as playing a role in the superconductivity \cite{ruf_strain-stabilized_2021}. The films grown by Ruf et al had room temperature resistivity of 58 $\mu \Omega cm$ which dropped to 13 $\mu \Omega cm$ at low temperatures, yielding a RRR of 4.5. On the other hand, they found that (101) oriented films show room temperature resistivity of 44 $\mu \Omega cm$ which dropped to around 1 $\mu \Omega cm$ at low temperatures, giving a RRR of 44, with no hints of superconductivity at the lowest temperature of 0.4 K. Both studies underscore the importance of substrate orientation, film thickness, and precise strain control achievable via MBE for accessing novel quantum phases, and together they position RuO$_2$ as a model system for strain-tunable superconductivity in simple binary oxides.

From the early attempts to grow thin films of RuO$_2$ mainly for technology-related applications \cite{frohlich_low-temperature_2003,jia_epitaxial_1996,carcia_electrical_1983}, methods such as CVD \cite{green_chemical_1985} and MOCVD showed room temperature resistivity of 30 $\mu\Omega cm$ at room temperature with the best RRR of around 30 \cite{frohlich_growth_2001}. The key challenges to grow of epitaxial RuO$_2$ films with low defects are similar to that of IrO$_2$. The lower lattice constant of RuO$_2$ (a = 4.49 $\textrm{\AA}$, c = 3.10 $\textrm{\AA}$) compared to TiO$_2$ substrates leads to significant strain, thereby defects, that affects the low temperature behavior. PLD \cite{tschirner_saturation_2023,kesler_epitaxial_2024} and MBE have been successfully used for obtaining the high quality epitaxial films to study the novel physics in this material. Traditional thermal effusion cells are ineffective for Ru metal source, so electron-beam evaporation is typically used \cite{uchida_superconductivity_2020,ruf_strain-stabilized_2021}. Developments in hybrid MBE using Ru(acac)$_3$, have also enabled high-quality films \cite{nunn_solid-source_2021} (see Figure \ref{fig:RuO2}). As mentioned in the section on IrO$_2$, these precursors are solid at room temperature and can be directly evaporated using an effusion cell operating around 100-200 $^{\circ}$C. Films grown with this approach had room temperature resistivities in the range of 330 $\mu\Omega cm$ for samples down to 0.8 nm and 168 $\mu\Omega cm$ for 9 nm (RRR = 1.2) \cite{rajapitamahuni_thickness-dependent_2024,jeong_metallicity_2025}, which compares well with previous reports with room temperature resistivity of 55 $\mu\Omega cm$ for a 15 nm film, close to the resistivity of bulk (RRR = 4) \cite{nunn_solid-source_2021}, and have provided additional insights into the anomalous Hall effect in this material. 

 In concluding this material, we must emphasize that there are still controversies surrounding RuO$_2$, especially with regards to altermagnetism.  One key aspect that must be considered is potential deviation among samples and the effect of interfaces and surface \cite{ho_symmetry-breaking_2025}. Another aspect needing resolution is the dichotomy between superconductivity and anomalous Hall effect in different films. Developing methods to understand each of these unique properties is needed to fully grasp the role of electronic interactions that could be harnessed for potential applications. Furthermore, combining superconductors with altermagnets is a topic that is being carefully scrutinized \cite{mazin_notes_2025} and has implications in the fundamental nature of superconducting states \cite{chakraborty_constraints_2024} and potential applications such as superconducting diodes \cite{banerjee_altermagnetic_2024} and josephson junctions \cite{papaj_andreev_2023}. RuO$_2$ would be an excellent candidate that can help uncover these details.

\subsubsection{Other rutiles}
CrO$_2$ and OsO$_2$ are the other rutiles that show ultra-high conductivity. CrO$_2$ has room temperature resistivity of 60 $\mu\Omega cm$ \cite{heffernan_role_2016} and has been long-studied in the context of being a half-metal \cite{Coey2002}. A half metal is a subclass of ferromagnets where the density of states at the Fermi level is solely composed of one spin orientation  which results in an effective metallic state for that spin but insulating state for the opposite spin \cite{de_groot_new_1983}. Bulk CrO$_2$ single crystals are difficult to grow due to the higher oxidation state required for Cr \cite{chamberland_crystal_1967}. More specifically, CrO$_2$ spontaneously converts to Cr$_2$O$_3$ when heated at atmospheric pressures. Success for thin film synthesis came through high-pressure methods such as CVD that utilized CrO$_3$ as a precursor. These reports have obtained films with room temperature resistivity of about 140 $\mu\Omega cm$ and RRR of around 60 \cite{watts_evidence_2000}. MBE growth, however, has proved significantly challenging with only reports of failures \cite{Ivanov2001,Ingle2001,rabe_preparation_2000}, which is linked to the difficulty in stabilizing this particular valence state. This was highlighted by Ingle et. al. who were able to stabilize both the lower valence state (Cr$_2$O$_3$) as well as Cr$^{6+}$ (CrO$_3$) using atomic oxygen via an RF plasma source. This indicates that the difficulty might not be a lack of oxidation power, but, rather, related to an instability of the kinetics and structural formation of CrO$_2$ in the rutile structure \cite{Ingle2001}. Going forward, it is an interesting open question regarding if recent advances in MBE technologies can be used to stabilize CrO$_2$ in the rutile structure, with, for example, metal organics or strain-stabiliztion, which may allow accessing different regimes that are sufficiently favorable for oxidation while having the correct kinetics to form the 4+ valence state.  

OsO$_2$ is the final member of this family and also has the lowest  resistivity  at 15 $\mu\Omega cm$ \cite{yen_growth_2004}. OsO$_2$ has been only synthesized in the bulk \cite{yen_growth_2004} and limited reports exist for the properties of this material, though there are theoritical predictions of novel properties that stem from the magnetism that is analogous to RuO$_2$ \cite{raghuvanshi_altermagnetic_2025} and strong spin-orbit coupling that can lead to large spin Hall effects \cite{zhang_strain-induced_2025}. There are several challenges with creating OsO$_2$, however. Os is a refractory metal and is also very rare, which makes it not only challenging for an MBE process but presents a financial barrier that stands in the way investigations. The most challenging problem, however, is the high level of toxicity of OsO$_4$, which are likely to be encountered during synthesis. Although there are significant challenges, there are many opportunities to explore the properties of OsO$_2$.

\subsection{High conductivity binary rocksalts}
There are only few examples of rocksalt binary oxides exhibiting high or ultra-high conductivity, with NbO, TiO and VO being the prominent materials. These systems are, however, not line-compounds and do not form in perfect stoichiometry. Rather, they tend to form with varying oxygen content and are typically represented as $A$O$_x$ where $x$ can be more or less than 1 by as much as 40 \% \cite{goodenough_anomalous_1971,hulm_superconductivity_1972}. Furthermore, this variation in stoichiometry has a large effect on their electrical properties \cite{goodenough_anomalous_1971,banus_electrical_1972,hulm_superconductivity_1972,chandrashekhar_electrical_1970}. NbO$_x$ is the most conductive among them, with resistivity of about 21 $\mu\Omega cm$ \cite{chandrashekhar_electrical_1970,hulm_superconductivity_1972}. NbO growth has been attempted by MBE on LiNbO$_3$ and Al$_2$O$_3$ substrates and it was found that polycrystalline films can be stabilized in low oxygen environment \cite{petrucci_growth_1988}. Sputter-deposited polycrystalline NbO$_x$ films were also tested as gate electrodes for Si based field effect devices \cite{gao_nbo_2004}. There are additional rocksalt and rocksalt-like suboxides of VO and TiO with a low resistivity around 100 $\mu\Omega cm$ \cite{hulm_superconductivity_1972}. Furthermore, TiO and VO are also superconductors with critical temperatures of about 0.5 K and 1.4 K, respectively \cite{hulm_superconductivity_1972}. Ultrathin VO film was grown on metallic Cu substrates namely as a buffer layer \cite{kishi_preparation_1996}. 

The unique aspect of off-stoichiometry and metastability of the metallic rocksalt phases have largely inhibited stabilization and application of epitaxial thin films. Yet, there are a few reports of films: for example, epitaxial TiO was obtained on MgO substrates using MBE with room temperature resistivity of about 300 $\mu\Omega cm$ and a superconducting transition at 0.45 K \cite{li_single-crystalline_2021}. However, the challenging aspect of off-stoichiometry in these materials, plus focus on the more stable forms, have inhibited research and understanding. As evidenced by recent report on stoichiometric TiO \cite{li_single-crystalline_2021}, it is likely that proper controls enabled by MBE may allow stabilizing high quality stoichiometric binary rocksalts that can further expand the understanding and application of these materials.

\subsection{High conductivity perovskites }
 The $AB$O$_3$ perovskite family and its derivatives get their name from the mineral perovskite (CaTiO$_3$), which was discovered nearly 200 years ago. This class of materials exhibits an extremely wide pallette of chemistries \cite{schlom_thin_2008}. Here, choices for the $A$ and $B$ cations spans nearly all the elements on the periodic table and also readily host organic molecules. Moreover, the anion is not confined to oxygen but can accommodate S, Se, F, Cl and Br. This wide range of elemental choices is available due to the simple cubic structural motif where the $A$-site cations site at the corners of the cubic unit cell, and the $B$-site cations are at the center and octahedrally coordinated by the oxygen atoms that sit at the center of the 6 faces (see Figure \ref{fig:structures}). This structure can be understood by simple geometric models for ionic structure \cite{george_fundamentals_2020}. Based on the relative ionic radii for the oxygen, the A, and the B site cations there is a simple geometric rule to predict the stability of the perovskite structure, called the Goldschmidt tolerance factor \cite{goldschmidt_gesetze_1926}. Moreover, the tolerance factor predicts subtle distortions away from the perfect cubic perovskite structure into tetragonal, hexagonal, orthorhombic and rhombohedral derivatives. The specific distortions of these derivatives can be simply understood by rotations of the corner connected octahedral units about the principle axis in the so-called Glazer notion \cite{glazer_simple_1975}, which provide a simple means to understand the wealth of possible structural motifs within this class. 
 
 The structural and chemical possibilities in the perovskites gives rise to an equally large number of functional properties. This spans metals, insulators, and semiconductors, to novel dielectrics such as piezo- and ferroelectrics, to correlated phases such as a huge range of magnets and high-temperature superconductors. Moreover, this class gives rise to multiple functional phases that are often coupled or in close proximity, such as multiferroic ferro-electrics and magnets, as well as metal-insulator transitions that are strongly coupled to magnetism \cite{schlom_thin_2008,goodenough_electronic_2004}. The origins of this large number of phases are intimately linked to the band filling and the resulting charge, orbital and spin character which are coupled through the bond geometry. As such, modifying these physical elements and the structure through epitaxy opens many exciting routes to tailor functional properties \cite{rondinelli_control_2012}, making the perovskites front-runners for oxide-based technology. 
 
 However, an important component for most electronic motif based on epitaxial oxides is a high-conductivity metal. The physics that determine if a perovskite is metallic and the resulting conductivity is similar to the binary oxides mentioned in the introduction. Here, few examples of high conductivity metals exist where $B$ is in the row of 3d transition metals with more example for the 4d and 5d. Specifically, CaVO$_3$, SrVO$_3$, and LaNiO$_3$ are the most prominent example of high-conductivity metals with $B$ being an element from the 3d row. For 4d, there is SrNbO$_3$, SrMoO$_3$ and SrRuO$_3$. Finally for 5d there is SrTaO$_3$, as well as perovskites with vacant A-sites (WO$_3$ and ReO$_3$) and the tungsten bronze phases. For the tungsten bronze phases an alkali metal occupies the $A$-site position giving for example A$_x$WO$_3$ where $x < 1$. For more details regarding the breadth of the science known in the perovskites, there are extensive studies  \cite{goodenough_electronic_2004} and epitaxial thin films of some conductive materials have provided some useful insights \cite{biswas_epitaxial_2023}. In the following, we will focus on select ultra-high conductivity perovskite materials.

\subsubsection{SrVO$_3$ and CaVO$_3$}

SrVO$_3$ and CaVO$_3$ both exhibit high conductivities at room temperature. CaVO$_3$ has room temperature resistivity of 20 $\mu \Omega cm$ \cite{jung_high-field_2002}. Similarly, SrVO$_3$ exhibits a room temperature resisitivity of 26 $\mu \Omega cm$ \cite{chamberland_alkaline-earth_1971} (polycrystalline samples), and the resistivity of recently grown single crystals was 600 $\mu \Omega cm$ \cite{berry_laser_2022}. Both these materials have been long studied in the context of electron-electron correlation effects. The simple filling of a single electron in the $t_{2g}$ level and the non-magnetic ground state make these materials attractive for studying the effects of electron-electron correlations. Moreover, the structural difference among SrVO$_3$ and CaVO$_3$ make an interesting system for comparison: the simple cubic structure for SrVO$_3$ exhibits a bond-angle of 180$^{\circ}$, which is in contrast to the lower bond angle of orthorhombic CaVO$_3$ (160$^\circ$) \cite{inoue_fermi_2002}, which should exhibit stronger correlation effects since the bandwidth is simply related to net correlation strength by $\cos^2{\theta}$ \cite{inoue_systematic_1995,pavarini_mott_2004}.  The correlation effects in these $d^1$ systems are sufficiently small to remain well-seperated from the Mott metal-insulator transition that is found in similar $d^2$ perovskites such as LaVO$_3$, which should have partially filled bands yet exhibits a large Mott gap. Additional details regarding the physics of electron correlations for perovskite vanadates in general can be found in Brahlek et al \cite{brahlek_opportunities_2017}. 

The moderate effects of electron-electron correlations in  SrVO$_3$ and CaVO$_3$ has driven a unique application as promising materials for transparent conductors \cite{Zhang2016}. Transparent conductors (more often called transparent conducting oxides) are critical components for a wide variety of applications from flat panel displays to photovoltaics. The key challenge associated with having a material that is both optically transparent and electrically conductive is the fact that these properties are mutually exclusive. This is because electrical conductivity requires free electrons to conduct electrical current but the free electrons simultaneously act to reflect light. Thus, the more conductive a material is, the less the material is optically transparent. Or, more precisely, the higher the carrier density, the smaller the wavelength window over which a material can be transparent since the free electrons will effectively screen light below the plasma frequency. The previous approach to address the challenges to make a good transparent conductor was to start with an optically transparent wide bandgap semiconductors, then dope it to increase the carrier concentration until it becomes metallic. However, considering the Drude picture, $\omega_p = \sqrt\frac{n e^2}{m^* \varepsilon}$ where $\omega_p$ is the plasma frequency, $m^*$ is the effective mass and $\varepsilon$ is the permittivity, the resulting carrier density $n$ must be sufficiently low such that the plasma frequency remains below the visible, thereby not causing reflection by free carriers. Moreover, the material must not have optical transitions near the visible part of the spectrum, which also contributes to optical loses. The industry standard that strikes the best balance is the wide bandgap material In$_2$O$_3$ which can be degenerately doped into a metallic phase by Sn (i.e., ITO) \cite{ellmer_past_2012}. 

The effect of electron correlations in high-conductivity metals like SrVO$_3$ and CaVO$_3$ allows for a different strategy to balance conductivity and transparency. Since these perovskites are good metals with 1 electron per unit cell, the plasma frequency should be in the visible and the anticipation is that the transparency should be low because $\omega_p \sim \sqrt\frac{n}{m^*}$. The effect of electron-electron correlations is to directly increase the effective mass of the charge carriers, or renormalize by a factor $Z_k$ so that the effective mass becomes $m^*=m_b^*/Z_k$, where $m_b^*$ is the general band mass, and $Z_k$ is unity for uncorrelated materials and less than one for correlated systems. This, in turn, directly reduces the screened plasma energies below the visible part of the spectrum. For example, $Z_k$ is around 0.55 for SrVO$_3$ and 0.47 for CaVO$_3$ \cite{paul_strain_2019}.  Moreover, since SrVO$_3$ and CaVO$_3$ are good metals the conductivities are significantly higher than those of ITO. These vanadates also have minimal (but non-zero) interband transitions within the visible range due to the energetic separation of the V 3d conduction band from both the filled oxygen 2p states and unoccupied 3d $e_g$ states. While their optical transmission is slightly compromised compared to ITO due to higher energy transitions near the blue part of the spectrum, the trade-off is mitigated by the superior electrical conductivity. Furthermore, the short electron mean free paths at room temperature make correlated metals well-suited for use in ultrathin films ($<$10 nm), where conductivity losses due to surface and grain boundary scattering are less severe. This allows vanadate films to reach competitive figures of merit even at film thicknesses as low as 10 nm, compared to ITO, which requires over 100 nm to achieve a sufficiently high conductivity. Importantly, correlated metals are typically composed of earth-abundant and low-cost elements, offering significant cost and manufacturing advantages over indium-based transparent conductors. These factors collectively position high conductivity correlated metals like SrVO$_3$ and CaVO$_3$ as well as additional metals mentioned throughout this review as strong candidates for next-generation electro-optical components.

\begin{figure}
    \centering
    \includegraphics[width=0.9\textwidth]{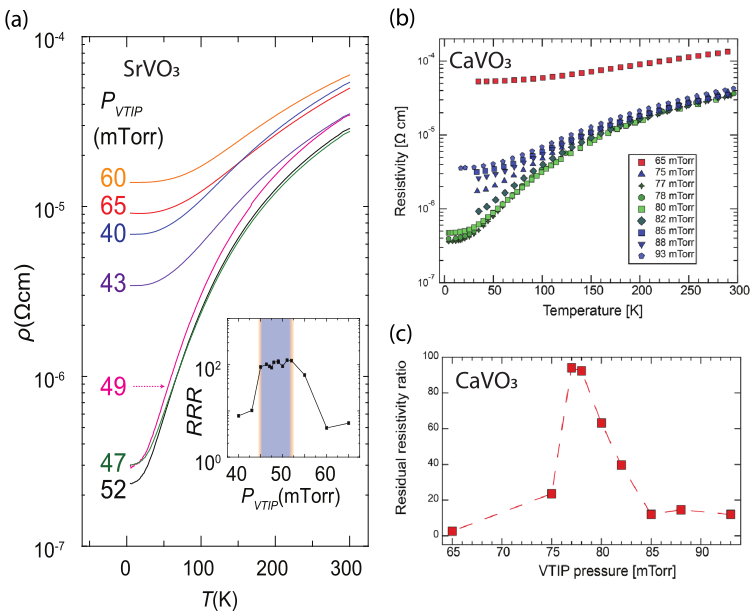}
    \caption{(a) Resistivity vs temperature graph for hybrid MBE growth of SrVO$_3$, with variations of metalorganic VTIP partial pressure \cite{brahlek_accessing_2015}. Reproduced from Brahlek, et.al., with the permission of AIP Publishing. Inset show RRR as a function of VTIP partial pressure, with highlights to indicate growth window. (b) Resistivity vs temperature graph for CaVO$_3$. (c) RRR vs VTIP partial pressure for CaVO$_3$. (b) and (c) are taken from from Kuznetsova et al \cite{kuznetsova_toward_2023}.}
    \label{fig:svocvo}
\end{figure}

There are two main challenges in growing phase pure SrVO$_3$ and CaVO$_3$ thin films. The first is stabilizing the vanadium in the correct valence and the second is controlling the cation stoichiometry. Regarding the valence: attaining the metallic phase requires a partially filled band at the Fermi level. For SrVO$_3$ and CaVO$_3$ with V$^{4+}$ results in d$^1$ filling, which is only attainable at highly reducing conditions, whereas V$^{5+}$ is favored at ambient conditions and higher pressures. This creates a challenge for epitaxial growth by techniques that utilize solid sources of Sr and V, i.e. PLD\cite{sheets_effect_2007,liberati_epitaxial_2009,cheikh_tuning_2024,marks_solid-phase_2021,mirjolet_high_2019}. The challenge is that these techniques require background gas to control plume kinetics prior to physisorption, typically of order 10 mTorr. This high oxygen pressure will tend to over oxidize the vanadium, yet lower pressure may result in high incident energies of incoming materials. Over-oxidation to V$^{5+}$ can result in insulating compounds such as Sr$_2$V$_2$O$_7$ and Sr$_3$V$_2$O$_8$ \cite{marks_solid-phase_2021}, as well as prominent surface oxidation post growth \cite{caspi_effect_2021}. MBE growth naturally takes place at a much lower partial pressures of oxygen with much lower kinetic energy, thereby giving more control. Oxide MBE utilizes thermal effusion cells for both Sr and V in an oxygen background, which enables accurate control to achieve V$^{4+}$ at pressures of 5$\times$10$^{-7}$ Torr of oxygen, thereby forming metallic SrVO$_3$\cite{shoham_scalable_2020}. 

Regarding challenge of cation stoichiometry control:  one of the highest reported RRR (about 12) for PLD grown SrVO$_3$ was performed with Ar as a background gas at a pressure, which was optimized around 150 mTorr \cite{mirjolet_high_2019}. RRR was reduced for lower and higher Ar pressure, whereas the growth rate reduced with increasing pressure. This behavior indicates the balance between controlling incident kinetics of the impinging atoms as well as net stoichiometry resulting from the sensitivity on the mass of the diffusion coefficients of Sr, V and O in the PLD plume. The growths of SrVO$_3$ by traditional oxide MBE relies on manual control over the Sr to V ratio which attained a bulk-like lattice parameter, indicating close to ideal stoichiometry. The maximum RRR of these films was 11 \cite{shoham_scalable_2020}, which possibly indicates defects due to limitations with in-situ flux calibrations. The hybrid MBE approach enables the use of the metal organic precursor vanadium oxitrisopropoxide (VTIP) as a vanadium source\cite{moyer_highly_2013, brahlek_accessing_2015,eaton_self-regulated_2017,kuznetsova_growth_2023} . The use of the precursor leverages two aspects to achieve higher quality materials. First, the V is directly bonded to oxygen which limits the necessity for a background of oxygen and seems to be able to stabilize V in not only the desired 4+ state for SrVO$_3$ but also for the more fragile 3+ for LaVO$_3$ \cite{brahlek_mapping_2016,brahlek_opportunities_2017,brahlek_accessing_2015,zhang_high-quality_2017}.  Moreover, the bonding to the organic ligands results in an energy barrier to incorporation that is tied to the supply of Sr, which results in a growth window that is quantifiable in both the structural properties (lattice parameter) as well as the electronic properties (RRR). This can be seen in Figure \ref{fig:svocvo}, where plots of resistivity versus temperature and RRR versus VTIP supply pressure both show an optimal "window" where the VTIP partial pressure can be varied yet resistivity and RRR do not change. However, over supply or under supply of VTIP does eventually change the stoichiometry of the film and results in a 10 times reduction of RRR as well as prominent structural defects\cite{eaton_self-regulated_2017}. The highest reported RRR values coincide with hybrid MBE growth window at over 200 \cite{moyer_highly_2013}, exceeding the highest for epitaxial films grown by other methods and bulk crystals \cite{chamberland_alkaline-earth_1971}. 

It is noted that the maximum measured RRR of 200 for hybrid MBE occurs in films grown on LSAT substrate, which puts SrVO$_3$ in slight tensile strain. The film thicknesses were around 50-60 nm, and going much thicker resulted in films relaxing and cracking. In general for thin film metals, the low temperature resistivity is quite sensitive to surface scattering \cite{elshabini‑riad_thin_1997,Kim2011} and the mean free path is limited to the order of the sample thickness. Therefore, it remains a question whether the RRR is due to the film geometry or is limited by the sample quality. Regarding the latter point, in hybrid MBE grown homoepitaxial La-doped SrTiO$_3$, RRR of 3000-5000 have been observed with mobilities higher than 30,000 cm$^2$V$^{-1}$s$^{-1}$. In contrast to the thin film SrVO$_3$, thickness of the La-doped layers were more than 10 times higher at 800 nm. As such, it remain an open question regarding the maximum RRR for SrVO$_3$, which could be significantly higher if samples could be grown thicker. Thus, the key open challenge is finding a substrate with near identical lattice match to SrVO$_3$ to enable thicker films. 

The high mobility of SrVO$_3$ has enabled finding new insights into the fundamental properties of the metallic phase of SrVO$_3$ that were hidden by defective samples \cite{brahlek_hidden_2024}. In the ultraclean limit, SrVO$_3$ displays anomalous transport behaviors that differ significantly from the simple picture of correlated electron, the Fermi liquid state. This is important to the broader understanding of correlations since SrVO$_3$ is one of the prototypical correlated materials that is used to benchmark first principle methods. Near room temperature the resistivity and the Hall effect exhibit temperature-dependent scaling that is reminiscent of strange-metal behavior observed in systems that display much stronger correlation effects, for example, the high-T$_c$ cuprates. At low temperatures (T$<$20 K) the resistivity seems to follow closer to the ideal Fermi-liquid behavior. Yet, application of high magnetic fields seems to highlight more nuanced physics. The Hall effect measurements suggest an apparent violation of Luttinger’s theorem, indicating that the carrier density inferred from transport deviates significantly from that predicted by band structure. These deviations cannot be explained by Fermi surface geometry, and instead point to a pronounced momentum-dependent transport relaxation times that varies across the different Fermi surface sheets. This anisotropic scattering, along with the identification of distinct transport regimes across temperature, highlights novelties in the electron-phonon or electron-electron scattering processes that were previously masked by dominance of defect scattering. The results underscore the importance of sample purity in accessing intrinsic correlated electron behavior and call for a reexamination of theoretical frameworks, such as DFT and DMFT, which accurately model disordered SrVO$_3$ but fail to capture the complexities unveiled in the ultraclean limit.

Finally, in contrast to SrVO$_3$, CaVO$_3$ presents additional challenges to achieving high quality samples. For SrVO$_3$, the key problems were controlling valence and cation stoichiometry, which is also true for CaVO$_3$, which has been largely mitigated through hybrid MBE growth \cite{eaton_self-regulated_2017}. This early work by Eaton et. al. showed that there exist a clear growth window where the Ca to V ratio is independent of flux ratio and the lattice parameter is close to the ideal value, similar to SrVO$_3$. However, the RRR was limited to a maximum of 30, compared to 200 for SrVO$_3$. It was thought that the unique challenge for CaVO$_3$ is that the structure is the orthorhombically distorted pervoskite due to the smaller A site ionic radius. Therefore, twins readily form which act as scattering centers limiting the RRR, which is in contrast to the case of simple cubic SrVO$_3$. However, the observed twins likely stem from the LaSrAlO$_4$ substrates used. Despite the good in-plane lattice match, LaSrAlO$_4$ substrates have distinct c-axis unit cells which creates vertical defects in CaVO$_3$. Revisiting the growth properties on LaAlO$_3$ showed dramatic improvements, yielding a maximum RRR near 100\cite{kuznetsova_toward_2023}. This highlights that although there are orthorhombic twins, mitigating the vertical defects due to the incommensurate c-axis unit cells does allow for cleaner materials. Lastly, moving away from rhombohedral LaAlO$_3$ and to a orthorhombic substrate could be a future direction for increasing the quality of CaVO$_3$. This is possible since the substrate structure and octahedral rotation pattern as well as strain combine to determine the octahedral tilt pattern and twin structure \cite{choquette_octahedral_2016,brahlek_structural_2017}. Given the larger V-O-V bond angle for CaVO$_3$, thus presumably larger electron-correlation effects, higher quality samples will enable contrasting the novelties of the electronic properties identified in SrVO$_3$ \cite{brahlek_hidden_2024}.

\subsubsection{SrNbO$_3$}

 From a basic electronic perspective SrNbO$_3$ is the 4d version of SrVO$_3$ with Nb exhibiting a nominal d$^1$ filling with a single electron occupying the lower lying $t_{2g}$ level. SrNbO$_3$ exhibits a similar room temperature conductivity to SrVO$_3$ for thin films at around 80 $\mu\Omega cm$ \cite {palakkal_off-stoichiometry_2025}, yet the much higher value of 330 $\mu\Omega cm$ in bulk \cite{xu_red_2012}. The large disparity of the properties of bulk SrNbO$_3$ relative to that of thin films highlights more complex structure and chemistry compared to its 3d cousin. Although SrNbO$_3$ was initially reported as cubic in the mid 1950s \cite{krylov_synthesis_1955,ridgley_preparation_1955},  the simple cubic perovskite structure is nearly unstable since the tolerance factor is close to 0.965, compared to 1.014 for cubic SrVO$_3$. This places SrNbO$_3$ close to the boundary of orthorhombic/tetragonal structure, similar to CaVO$_3$ (0.978) which is accepted to be orthorhombic. This is consistent with temperature dependence of the structural phase for SrNbO$_3$ where the room temperature structure is orthorhombic which gives way to tetragonal and cubic phases at high temperatures \cite{macquart_primitive_2010}. A second key point for a variety of reports is that attaining stoichiometric samples in the bulk is challenging in that samples readily form as Sr$_x$NbO$_3$, with reports of $x$ ranging from 0.75 to 1 \cite{isawa_photoelectron_1994,isawa_synthesis_1993,xu_red_2012,peng_synthesis_1998,istomin_perovskite-type_1998,hessen_crystallization_1991,isawa_hall_1993,hannerz_transmission_1999,macquart_primitive_2010,kodera_investigation_2018,seo_photoelectrochemical_2016}. Electronically, the Sr vacancies can be accommodated by the addition of an oxygen vacancy or by shifting 2 Nb atoms into the energetically preferred 5+ state \cite {roth_sputtered_2020,isawa_photoelectron_1994}. The psuedocubic lattice parameter ranges from around 3.99 $\textrm{\AA}$ for $x=0.75$ towards 4.015 - 4.02 $\textrm{\AA}$ as $x$ $\rightarrow$ 1 \cite{isawa_synthesis_1993}. This highlights the tolerance of perovskite structure to accommodate large density of vacancies through only moderate distortions, and draws interesting comparison of Sr$_x$NbO$_3$ to the tungsten bronze phases which coexists with the perovskite for  $x<$ 0.7\cite{isawa_synthesis_1993}.  
 
 Like SrVO$_3$, SrNbO$_3$ has seen a lot of interest due to the simultaneous high conductivity and optical transparency, making them excellent candidates for semi-transparent electronics \cite{roth_sputtered_2020,ha_design_2021,park_srnbo3_2020,jeong_transparent_2024,liu_all-inorganic_2024,mirjolet_optical_2021}. The key advantage of Sr$_x$NbO$_3$ is that it remains optically transparent for smaller wavelengths compared to SrVO$_3$. The screened plasmon energy for SrVO$_3$ 1.33 eV compared to the large value of 1.98 eV for Sr$_x$NbO$_3$, which pushes the absorption of light into the IR regimes and closer to the visible \cite{Zhang2016,park_srnbo3_2020}. This larger plasma frequency is driven by the weaker correlation effects in SrNbO$_3$ compared to that of the vanadates , which, due to the smaller renormalized mass (due to the larger renormalization factor, $Z_k$=0.72) \cite{paul_strain_2019}, should reduce the room temperature resistivity. However, the naturally higher level of disorder actually increases the room temperature resistivity for Sr$_x$NbO$_3$. The biggest gain in performance as a transparent conductor comes from the reduced interband transition at lower wavelengths into UV part of the spectrum\cite{park_srnbo3_2020}. Specifically, the transmission coefficient for Sr$_x$NbO$_3$ is above 0.80 for the entire visible part of the spectrum and remains above 0.60 for the entire UV portion. In comparison, the transmission coefficient for SrVO$_3$ drops below 0.80 near 2.5 eV (blue) down to 0.30 near 3 eV (violet), which confines the competitiveness of SrVO$_3$ as a transparent conductor to thin samples ($<$10 nm). Combining this with the high intrinsic conductivity that results from the large carrier density makes Sr$_x$NbO$_3$ competitive with the industry standard ITO in the visible and the UV for film thickness up to 100 nm.

All oxide electronics rely on the ability to interface semiconducting oxides with metals and insulators, which dictates how oxide metals can be used to make Ohmic contacts to semiconductors or as gate metals in contact with insulator where a Schottky contact may be desired. These questions delineate the necessity for understanding and utilizing interfaces between metals like Sr$_x$NbO$_3$ or SrVO$_3$ and semiconductors like SrTiO$_3$ that can be readily doped. For binary and compound seminconductors, these properties can be understood with a few simple set of rules known as the Anderson and the Schottky–Mott rules for creating band alignment diagrams. These enable understanding how charges redistribute at semiconductor interfaces to achieve electrostatic equilibrium.  However, these rules are found to be insufficient for oxides \cite{zhong_band_2017}. This inadequacy is due to the dominant role of factors such as strong electron correlation, orbital hybridization, and interface-specific structural distortions. In this context, a band alignment framework based on the alignment of oxygen 2p energy levels offers a more accurate approach to predicting charge transfer at oxide interfaces. First principle calculations were used to predict interface properties for the high conductivity metals SrVO$_3$, SrNbO$_3$ and SrTaO$_3$ with the semicondutor SrTiO$_3$ and demonstrate the validity of this approach: SrVO$_3$ forms a Schottky barrier with negligible interfacial conduction, whereas SrNbO$_3$ and SrTaO$_3$ form distinct Ohmic contacts characterized by substantial charge transfer and the formation of parallel conduction in the SrTiO$_3$\cite{zhang_design_2022}. These predictions are supported by experimental data for Sr(V,Nb,Ta)O$_3$ films that were grown directly on SrTiO$_3$ as well as insulating substrates using PLD in high vacuum and temperatures around 650 $^{\circ}$C \cite{zhang_design_2022}. In vacuum at these elevated temperatures SrTiO$_3$ readily losses oxygen and will become metallic and exhibit high mobilities and RRR. The SrVO$_3$/SrTiO$_3$ systems showed RRR of order of 1, which is typical for PLD grown materials. This directly showed that any excess charge in the SrTiO$_3$ was electrostatically transferred into the SrVO$_3$, thus confirming the predicted Schottky junction. On the other hand, for Sr$_x$NbO$_3$ and SrTaO$_3$, theory predicts the opposite charge transfer and an Ohmic junction. This is borne out in experiments where RRR is of order of 10,000 and high mobility. Control samples grown on non-conductive substrates exhibited RRR around unity that is intrinsic to the film \cite{zhang_design_2022,oka_intrinsic_2015}. Moreover, magnetoresistance and Hall effect showed clearly that there are two conductive channels, which are found to be consistent with a high-mobility SrTiO$_3$ conduction channel and a thin, low mobility metal \cite{zhang_extremely_2021,zhang_design_2022,wei_high-mobility_2024}. When taken with direct $I-V$ measurement, this clearly shows that charge is transferred into the SrTiO$_3$. Interestingly, this theory provides a simple explanation for a range of unusual transport phenomena that have been attributed to more complex topological physics \cite{mohanta_semi-dirac_2021,ok_correlated_2021}. These results establish oxygen 2p alignment as a predictive metric for interface design and highlight the broader potential of metallic perovskites not only as low-resistance contacts but also as active components in oxide electronics. By enabling controlled access to interfacial states and tailored charge transfer, this framework opens new pathways for designing oxide-based devices that exploit correlated transport, electrostatic modulation, and emergent quantum phenomena.

In analogy to the vanadates, there are similar challenges to achieving high-quality SrNbO$_3$ films. The first is stabilizing the correct valence state since 5+ state is preferred over the desired 4+ state. Second is controlling the correct Sr to Nb ratio. However, unlike vanadium, niobium is refractory and thus cannot be used in a standard MBE process, thus there are few reports for MBE-grown Sr$_x$NbO$_3$. Regarding the first challenge, for PLD, growth pressures have mitigated the formation of Nb$^{5+}$. For example, growth in vacuum of $<$10$^{-8}$ Torr have yielded high conductivity films with resistivity of 28 $\mu\Omega cm$ \cite{oka_intrinsic_2015}. Interestingly, films have been grown at higher oxygen pressures of $\sim $10$^{-5}$ Torr  and detailed spectroscopy measurements showed clear band structure that was consistent with first principle calculations and showed that the measured Fermi surface yields a carrier density close to the expected single electron per unit cell for Sr$_x$NbO$_3$ \cite{bigi2020direct}. Yet, valence state measurements showed that the surface contains both the desired Nb$^{4+}$ as well as Nb$^{5+}$, which highlights the extreme sensitivity to oxygen. Sputtering has been shown to be a viable route to yield films with high conductivity with room temperature resistivity around 320 $\mu\Omega cm$ \cite{roth_sputtered_2020}. The films were grown in an RF high vacuum sputter system on LSAT with background pressure of 20 mTorr if Ar, and film temperature was 700 $^{\circ}$C. Interestingly, even though the target used for the growth was stoichiometric Sr$_2$Nb$_2$O$_7$, the resulting Sr$_x$NbO$_3$ films were off stoichiometric with the $x$ ranging from 0.63 to 0.78. However, the overall structural, optical and transport properties indicate good quality films with high figures of merit as transparent conductors.  

MBE remains a promising yet underexplored technique for SrNbO$_3$ with only a few reports from recent years \cite{mahatara_high_2022,thapa_surface_2022,palakkal_growth_2024, palakkal_off-stoichiometry_2025}. Given the challenge with close to unity cation stoichiometry in both bulk samples as well as sputtered and PLD grown films, the use of independent control over Sr and Nb in a low oxygen background could allow for precise control over cation stoichiometry and Nb valence. The central challenge, however, is supplying Nb since it is refractory and cannot be used in a traditional MBE cell. Rather, electron-beam evaporation is required. Palakkal et al recently reported growth of Sr$_x$NbO$_3$ by MBE, which utilized electron-beam for evaporating Nb (and also Sr) in oxygen background and a variety of substrates\cite{palakkal_off-stoichiometry_2025}. This work showed clear stabilization of the Sr$_x$NbO$_3$ structure with room temperature resistivity ranging from 800 $\mu\Omega cm$ down to as low as 30 $\mu\Omega cm$ , although the low RRR of 1.4 suggests that the low temperature properties are severely limited by defects in the films and chemical reactions with the substrate \cite{palakkal_growth_2024}. Given the success of hybrid MBE in stabilizing V$^{4+}$ in SrVO$_3$ as well as a host of other materials, an analogous Nb-containing precursor may enable a similarly well-defined growth window for SrNbO$_3$. However, the lack of a volatile, well-behaved Nb precursor for hybrid MBE remains a bottleneck as well as questions regarding the process for incorporation Sr and Nb into the film and if this can result in close to stoichiometric materials. Recent work employing tris(diethylamido)(tert-butylimido) niobium (TBTDEN) as a Nb source have achieved promising results \cite{thapa_surface_2022,mahatara_high_2022}. This foray in hybrid MBE for this high conductivity metal has resulted in films of high structural quality, but highlight the key difficulty of controlling the valence in the desired 4+ state. In situ x-ray photoemission spectroscopy showed the high sensitivity of the surfaces to the formation of Nb$^{5+}$, which spontaneously formed post growth. This was consistent with previous reports for PLD growth which showed a mixed surface of  4+ and 5+ states \cite{bigi2020direct}. The authors showed that this highly sensitive surface can be readily capped with protective and functional layers - either SrHfO$_3$ or BaSnO$_3$. Although there are clear challenges, this shows that the field of MBE growth of Sr$_x$NbO$_3$ is wide open and there are clear interesting studies remaining regarding the growth by MBE as well as promise as a functional oxide element.

\begin{figure}
    \centering
    \includegraphics[width=0.8\textwidth]{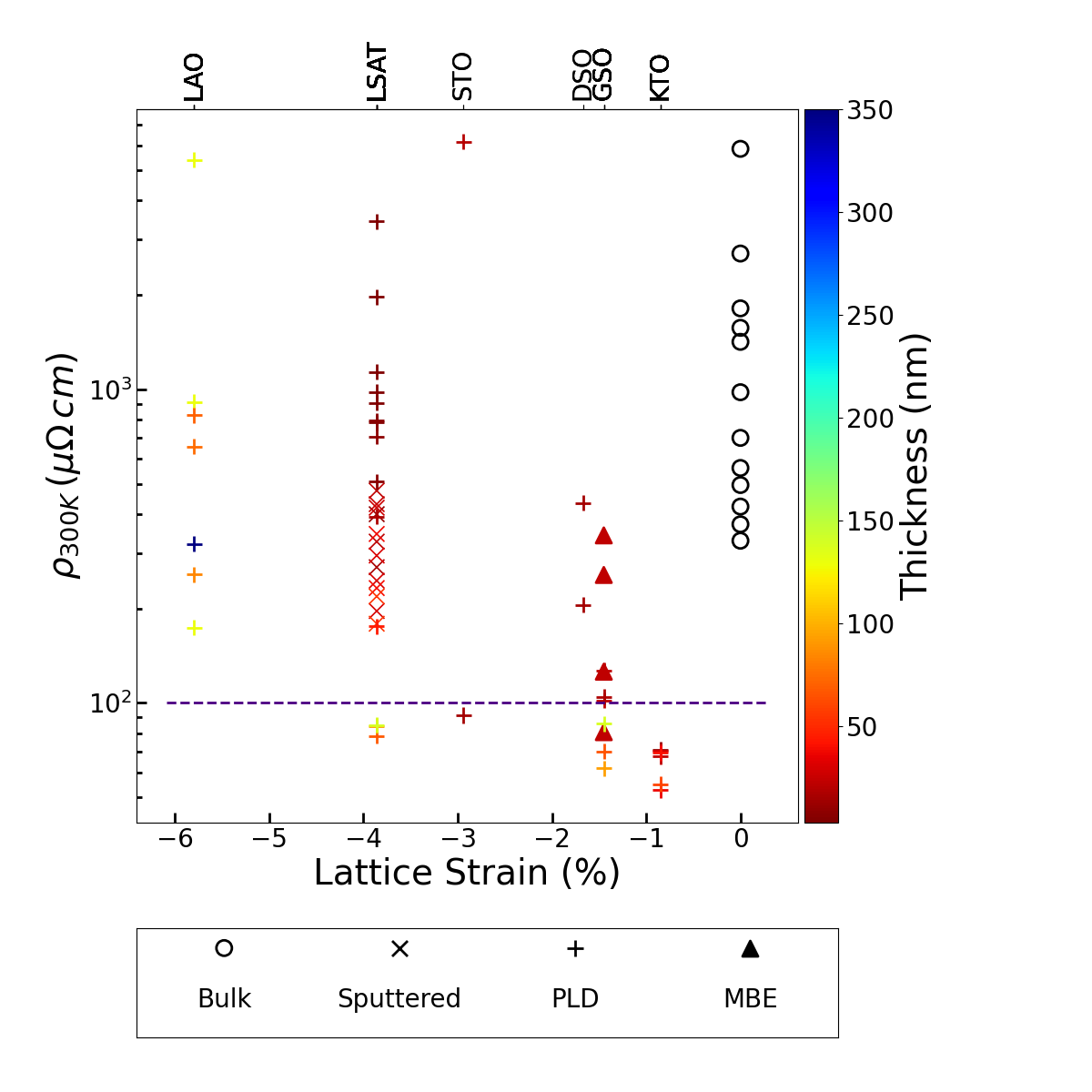}
    \caption{ Strain dependence of room temperature resistivity and RRR in SrNbO$_3$ films grown by different techniques. The substrates corresponding to the relative strain are indicated: LAO=LaAlO$_3$, LSAT=(LaAlO$_3$)$_{0.3}$(Sr$_2$AlTaO$_6$)$_{0.7}$, STO=SrTiO$_3$, DSO=DyScO$_3$, GSO=GdScO$_3$, KTO=KTaO$_3$.}
    \label{fig:SrNbO3}
\end{figure}

\subsubsection{SrMoO$_3$}
SrMoO$_3$ is the perovskite oxide with the lowest room temperature resistivity at 5 $\mu\Omega cm$ which was reported for bulk crystals grown using floating zone technique under extreme reducing conditions \cite{nagai_highest_2005}. SrMoO$_3$ crystallizes in a cubic perovskite structure similar to SrVO$_3$ and considering the valence of Mo$^{4+}$ gives nominal filling of two electrons in the $t_{2g}$ level. SrMoO$_3$ has arisen as a promising transparent conductive oxide materials which is due to the non-negligable correlations effects and low room temperature resistivity \cite{jia_epitaxial_1996, radetinac_optical_2016, cappelli_electronic_2022}. The correlation strength of SrMoO$_3$ is a relatively small amount of renormalized band \cite{paul_strain_2019}. However, there are predictions that correlation effects do arise: SrMoO$_3$ is predicted to be a Hund's metal where the Hund's coupling is responsible for band renormalization rather than onsite repulsion \cite{Georges2013}. This results in an effective renormalization similar to the vanadates and niobates around 0.7 and a screened plasma frequency of 1.7 eV \cite{mizoguchi_optical_2000}. This prediction is in good agreement with experimental results from PLD grown SrMoO$_3$ that was in-situ transferred for ARPES and gave a mass renormalization of $Z_k$ = 0.58 \cite{cappelli_electronic_2022}. Moreover, the optical transmission coefficient throughout the visible to ultraviolet ranges (500 to 300 nm) remains above 0.60-0.70 which is competitive with Sr$_x$NbO$_3$\cite{kuznetsova_growth_2023}. The combination of high optical transmission and the ultrahigh conductivity makes SrMoO$_3$ a competitive transparent conductive material. Finally, there has been interesting work on utilization of the high conductivity of SrMoO$_3$ as an active element for all-oxide RF applications \cite{radetinac_highly_2014}. In previous experiments, Pt has been integrated as an epitaxial bottom electrode for Ba$_x$Sr$_{1-x}$Ti$_x$O$_3$ resonators \cite{mikheev_electric_2012}. However, epitaxial Pt is of low quality and will tend to reorient near the growth temperatures of oxides. These structural defects as well as additional non-idealities that arise at heterointerfaces compromises the overall quality. Since the resistivity of SrMoO$_3$ is close to that of the noble metals and a close lattice match to dielectrics such as BaTiO$_3$ make it an excellent candidate for future work as a bottom electrode in RF applications.

The key problems with synthesizing SrMoO$_3$ is that Mo readily takes on higher valence states in the SrMoO$_4$ structure with Mo being in the 6+ state. Therefore highly reducing conditions are necessary. Nagai et. al. reported that oxygen partial of 10$^{-25}$ atmospheres was necessary in a specially designed floatzone system \cite{nagai_highest_2005}. SrMoO$_3$ thin films have been synthesized using various techniques including RF magnetron sputtering \cite{mizoguchi_optical_2000}, pulsed laser deposition (PLD) \cite{radetinac_optical_2016,cappelli_electronic_2022,wang_experimental_2024}, and MBE \cite{kuznetsova_growth_2023,takatsu_molecular_2020}. There are several main challenge for the growth. As mentioned above there is the extreme sensitivity to the oxygen pressure to avoid over oxidation of Mo. There are two additional challenges for thin films. The first is the difficulty in evaporating Mo, which  requires electron-beam evaporation for an MBE process, or utilizing a solid source technique like sputtering or PLD. The remaining challenge is due to the natural volatility of the oxides of Mo, which may lead to non-stoichiometric materials. PLD growth has lead to films with resistivities as low as 20-30 $\mu\Omega cm$\cite{alff_role_2013,radetinac_optical_2016}. This was achieved using near perfect lattice matched GdScO$_3$ (0.2 percent) grown at 650 $^{\circ}$C with a thin SrTiO$_3$ buffer layer. The challenge of oxygen was mitigated through the use of a mixed background gas during PLD growth at 50 mTorr. The background gas was composed of Ar and 2.5 percent H$_2$, which act to reduce any residual oxygen. Additionally, NO was similarly proposed. It was noted that with pure Ar the minimum growth temperature was 700 $^{\circ}$C and below this Mo tended to be over-oxidized. Interestingly, the films grown at slightly lower temperature with the addition of small portion of H$_2$ had superior quality in that they had resistivity of 20 $\mu\Omega cm$ as well as exhibiting clean XPS peaks indicating Mo$^{4+}$ without any signature of metallic Mo and higher structural quality \cite{alff_role_2013}. 

MBE based growth methods have initially focused on combining a conventional effusion cell for Sr and an electron-beam evaporator for Mo \cite{takatsu_molecular_2020}. The substrates were SrTiO$_3$ and KTaO$_3$ and growth temperatures optimized at 520 $^{\circ}$C and a thin SrTiO$_3$ buffer layer. The oxygen pressure was keep at 4$\times$10$^{-7}$ Torr during growth. This resulted in films with high structural quality as well as room temperature resistivity of 117 $\mu\Omega cm$ without the SrTiO$_3$ buffer layer and 34 $\mu\Omega cm$ with the buffer layer. A small RRR of 1.7-1.8 implies that the films still contain significant defects. The likely source of defects could be either cation non-stoichiometry or the presence of a small amount of Mo of different valence that result from over oxidation with the background gas or from diffusion from the SrTiO$_3$ interface, with the latter likely as highlighted by in-situ XPS. Moreover, these films showed clean band structure in good agreement with theory calculations.  

To mitigate difficulties associated with the use of electron-beam evaporation of Mo, hybrid MBE approach has been considered. However, no metal-organic precursors were found in high enough purity to enable this approach. Rather, direct evaporation of the suboxides of Mo have shown considerable success. In this approach solid Sr is used in conjunction with MoO$_3$, both in standard MBE cells\cite{kuznetsova_growth_2023}. This leverages the high volatility of the oxides compared to the metal. However, temperature (540-560 $^{\circ}$C) required to produce the necessary vapor pressure that is required to get a sufficient growth rate (flux of 1$\times$10$^{13}$ atoms/cm$^{-2}$s$^{-1}$) is close to the point where MoO$_3$ reduces to a metal and gaseous oxygen. This poisoning of the source was observed by a clear and systematic dropping of the flux over time in chamber pressures of 2.0$\times$10$^{-8}$ Torr. To mitigate this, Kuznetsova et. al. modified their MBE reactor to direct atomic oxygen directly into the mount of the crucible \cite{kuznetsova_growth_2023}. Flux measurements showed stability in oxygen pressure of 4.5$\times$10$^{-6}$ Torr for 2 hours with this modification. The optimum growth conditions were found to be 1.5$\times$10$^{-7}$ Torr oxygen pressure and a temperature of 700 $^{\circ}$C on LSAT substrates. The lattice parameter for all films was lower than that of bulk (3.975 $\mathrm{\AA}$ versus 4.054 $\mathrm{\AA}$), possibly due to residual strain. Interestingly, systematic stoichiometry dependence showed that substoichiometric films showed clear SrMoO$_4$ peaks. Moreover, the lattice parameter showed a systematic increase with increase in MoO$_3$ flux and an overall step-like structure. Although the origin of this behavior remains unknown it may be related to strain relaxation effects combined with how the structure accommodates non-stoichiometric defects. Finally, the optimized films resulted in resistivity as low as 50 $\mu\Omega cm$, which, when combined detailed optical measurements, placed them on par with Sr$_x$NbO$_3$. Taken together, these results indicate that it is possible to grow high quality thin films of SrMoO$_3$, yet there is much room for additional improvements and studies for a wide array of applications.

\subsubsection{Tungsten Bronze}
Tungsten bronze, represented as $A_x$WO$_3$, are related to the perovskite structure in that they are structurally composed of $B$O$_6$ octahedral unit that are corner connected with $A$-site cations at the corners of the pseudocubic unit cell. The key difference is that there are vacancies at the $A$-sites. The typical $A$-site is an alkali metal and $x < 1$ and represents a large class of materials \cite{dickens_tungsten_1968}. The Sr$_x$NbO$_3$ system discussed earlier exhibits tungsten bronze phase for $x$ around 0.60. The properties that these materials exhibit range from insulator and semiconductors to high conductivity metals and superconductors. The high conductivity metals have resistivity on the order of 100 $\mu\Omega cm$. WO$_3$ and variants are electrochromic and superconducting \cite{raub_superconductivity_1964,haldolaarachchige_superconducting_2014,soma_epitaxial_2016} and there are reports of possible high temperature surface superconductivity about 90 K \cite{reich_possible_1999}. 

This class of material is particularly challenging for thin film synthesis mainly due to the volatility of the $A$-site and the tendency for oxygen vacancy formation, as well as the refractory nature of the B species (namely tungsten). Epitaxial films have mostly been grown using PLD \cite{wu_synthesis_2014,mattoni_charge_2018,soma_epitaxial_2016} and sputtering \cite{leng_insulator_2017} and the conditions under which the films are realized make this class of material inherently challenging to grow. There is a particular challenge for MBE growth: first the low vapor pressure of W can only be circumvented using electron-beam evaporation or hybrid approach. That, combined with the high vapor pressure of the alkali metals, makes the growth even more challenging. However the high vapor pressure of the alkali metals and the low sublimation points of their oxides provide opportunities for adsorption controlled and self-limited growth. We also point out potential cross contamination issues that may arise due to the volatility of the alkali metal sources in MBE growth reactors. Interestingly, novel approaches for controlling the sources have proven successful. For example, alloying the alkali metal with another metal that has both a low melting point and low vapor pressure such as Sn with Li \cite{du_control_2020} and In with Cs \cite{parzyck_single-crystal_2022}. Creating the alloy can be readily done in a glove box then transported safely in the air. The mismatch in vapor pressures then lets the alkali metal to directly evaporate out of the mixture in the MBE in a highly controllable process. Altogether the tungsten bronze family of materials exhibit members with high conductivities and is epitaxially matched to a wide array and is not deeply explored as thin films. As such this is a vast space for future materials innovations.

\subsubsection{Other perovskites and perovskite-like materials}

ReO$_3$ is a perovskite-type material with high conductivity of about 9.2 $\mu\Omega cm$ at room temperature \cite{quirk_singular_2021}. High quality bulk crystals have been obtained with RRR on the order of 1500, but films have been difficult to obtain and MBE-grown films do not exist, presenting opportunities for understanding the synthesis science and effects of heterostructures and interfaces. A main challenge is the low vapor pressure of Re, which could be circumvented by electron-beam evaporation, and hybrid MBE if suitable precursors can be identified.  

LaNiO$_3$ is another highly conductive perovskite that falls under a class of strongly correlated nickel-based materials which is a precursor to the superconducting infinite-layer nickelates. LaNiO$_3$ is the only member of the rare-earth nickelates that stabilizes in the rombohedral $R\bar{3}c$ structure, but can be grown epitaxially in a pseudocubic form by choosing proper substrates \cite{Catalan2008}. Ni adopts $t_{2g}^6 e_g^1$ ($d^7$) state and the NiO$_6$ octahedron exhibits rotations which were well-quantified in MBE-grown films \cite{may_quantifying_2010}. Bulk crystal synthesis of nickelates were challenging because of the higher oxidation state needed for Ni, and the only way to stabilize the Ni$^{3+}$ state is by applying high pressure during synthesis. However, epitaxial films can be readily stabilized. Extensive research on the rare-earth nickelates has provided insights into their properties and the physics of correlated systems, and their potential for applications in devices such as memristors. There are some excellent reviews that focus on LaNiO$_3$ and other nickelates, including those on epitaxial films and heterostructures \cite{medarde_structural_1997,Catalan2008,middey_physics_2016,catalano_rare-earth_2018}.

Among major achievements in nickelates, the observation of superconductivity is noteworthy. The structure of the infinite-layer nickelate compounds were precicted to be similar to superconducting cuprates but stabilizing the required Ni$^{+1}$ state proved to be much challenging. Via a clever topotactic reduction reaction of epitaxial perovskite nickelates, superconductivity was finally discovered in the infinite layer NdNiO$_2$ \cite{Li2019_superconductivity_in_nickelate}. Following that, numerous types of infinite-layer superconducting nickelates were discovered \cite{pan_superconductivity_2021,sun_electronic_2025}, including LaNiO$_3$ based ones grown using MBE \cite{sun_electronic_2025}. A noteworthy finding is the observation of superconductivity by topotatic reduction of MBE-grown n=5 Ruddlesden-Popper phase (Nd$_6$Ni$_5$O$_{16}$) to the infinite-layer Nd$_6$Ni$_5$O$_{12}$, Ni$^{+1.2}$ state in which an optimal $d^{8.8}$ band filling was achieved without the need for extrinsic doping \cite{pan_superconductivity_2021}. The highest critical temperature reported so far on these reduced films is around 40 K in a complex nickelate with multiple cations \cite{chow_bulk_2025}. These discoveries led to renewed efforts in bulk crystals of Ruddlesden-Popper phases which allowed the discovery of superconductivity in the n=2 phase of La$_3$Ni$_2$O$_7$ up to around 80 K under high pressures \cite{sun_signatures_2023,zhang_high-temperature_2024}. As a result, strained epitaxial films have a large potential to further increase the superconducting transition temperatures. Due to their similarity to the superconducting cuprates, electronic phases such as charge density waves and strange metals have also been observed \cite{wang_experimental_2024}. The study of superconducting nickelates is still a nascent field and much research is being carried out in order to discern the similarities and differences with the superconducting cuprates and understand the mechanisms of high-temperature superconductivity. There are some recent reviews on these infinite-layer compounds that pertain to the synthesis science and electronic behaviors \cite{lee_aspects_2020,chow_infinite-layer_2022,wang_experimental_2024}. The potential incorporation of LaNiO$_3$ into devices exploiting these novel phenomena provides opportunities for quantum-based applications.

\subsection{Metallic delafossites }
The metallic delafossites are a relatively understudied class of materials compared to the perovksites and binary oxides. These materials have the general chemical formula $AB$O$_2$ where $A$ is a monovalent transition metal cation and $B$ is a trivalent (transition metal or group 3) cation \cite{Marquardt2006}. The delafossites have unique layered structure composed of units of $A$ and $B$O$_2$ which repeat along the c-axis, as shown in Figure \ref{fig:structures}. The $A$-layers are configured with direct A-A bonding in the $a-b$ plane and then bonded vertically to oxygen. Within the $B$O$_2$ layers, the $B$-sites are in octahedral configuration which are edge-shared. The in-plane lattice is triangular and the overall crystal symmetry is in $R\bar{3}m$. Specific members of this group, namely PtCoO$_2$, PdCoO$_2$, PdCrO$_2$ and PdRhO$_2$ are the most conductive oxides known and rival that of noble metals with room temperature resistivity as low as 2 $\mu\Omega cm$ for PdCoO$_2$\cite{Mackenzie2017, Hicks2012}. Although understudied \cite{Mackenzie2017}, the first bulk single crystals were synthesized many decades ago at Dupont \cite{Shannon1971,Prewitt1971,Rogers1971}. Interestingly, the Pt compounds were formed accidentally through hydrothermal reaction of Co$_3$O$_4$ with the Pt reaction tube in an HCl solution \cite{Shannon1971}. These early works noted several interesting facts including the ultrahigh conductivity at room temperature with RRR in excess of 100, as well as the novel chemistry with Pd and Pt in 1+ state. Following this early work, recent attempts in the 1990s through the 2000s have allowed these materials to be realized with high quality and exceedingly low impurity levels and RRR values in excess of 400  \cite{tanaka_growth_1996,tanaka_crystal_1997,Takatsu2007,Hicks2012}.  

In general, the metallic nature in delafossites is directly linked to the unique bonding geometry. Specifically, the direct metallic bonds within the $a-b$ plane is unique to oxides and mimics the bonding in metals as well as their high density of states at the Fermi level. The bonding in the a-b plane is thus responsible for the high conductivity, yet the natural structural layering and chemical properties manifest in strong anisotropy in the overall electronic structure. The metallic $A$ layers are separated along the c-axis by $B$O$_2$ layers, which exhibit properties more traditionally associated with oxides. Within these layers the $B$-site is in a 3+ state, which, considering the example of Co is in a $d^6$ state. In octahedral crystal field, Co$^{3+}$ has a configuration with a completely filled $t_{2g}$ level and the higher lying $e_g$ is completely empty, and thus is highly insulating. The spatial separation and high energy barrier to hopping between the metallic $A$ layers results a very low conductivity along the c-axis relative to the conductivity in the $a-b$ plane. More specifically, the resistivity along $a-b$ plane is more than 3 orders of magnitude higher than that along c-axis\cite{takatsu_anisotropy_2010,Hicks2012,Daou2015}. The properties of the electrons within the $A$ layers are effectively 2D with almost no dispersion along the c-axis. 

The key question then arises: why is the resistivity so low for the delafossites? Are the electronic properties insensitive to defects or do the materials just grow with anomalously low density of defects? This question was answered by intentionally introducing defects into the delafossite structure: tracking how the resistivity changes with dose implies that the pristine state contains very few defects \cite{Sunko2020}. The $A$ layers are also resistant to defects and damage, and defects are mainly localized in the $B$O$_2$ layer \cite{zhang_crystal-chemical_2024}. The role of electron phonon coupling has also been highlighted in the high conductivity behavior of PdCoO$_2$ \cite{homes_perfect_2019,seo_interaction_2023}. This leads to extremely high mean free path and effectively gives rise to such high conductivity values. Due to long mean free paths (tens of $\mu m$), the discovery of viscous nature of electrons and hydrodynamic electronic transport was possible \cite{Moll2016}. Although the metallicity is determined by the A layer, the BO$_2$ layer is insulating and can be strongly correlated which leads to novel properties and emergent behaviors. For example, in the case of PdCoO$_2$, Co$^{3+}$ exists in low spin state, and the material is a typical Pauli paramagnet. On the other hand, in PdCrO$_2$ the Cr$^{3+}$ state leads to antiferromagnetic order, and evidence of Kondo type interactions have been observed between the conduction electrons and the insulating layer \cite{Sunko2018}. Furthermore, novel phenomena such as quantized magnetoresistance oscillations \cite{Putzke2020}, surface and emergent magnetism \cite{Mazzola2017,Harada2020,brahlek_emergent_2023} and Planckian behavior \cite{zhakina_investigation_2023} are observed. Another interesting aspect of the layered nature of the delafossites is the polar nature of the surface, which can result in interesting physics. For example, in PdCoO$_2$, the different terminations (Pd or CoO$_2$) can result in different types of surface states \cite{Mazzola2017,Harada2020,kong_fully_2022,song_surface_2024} which are largely affected by disorder \cite{Mazzola2022}. As a result, controlling the termination through layer-by-layer growth provides a way to control the surface behavior. The main properties of these materials can be found in another review that focuses on bulk single crystals \cite{Mackenzie2017}.

\begin{figure}
    \centering
    \includegraphics[width=0.9\textwidth]{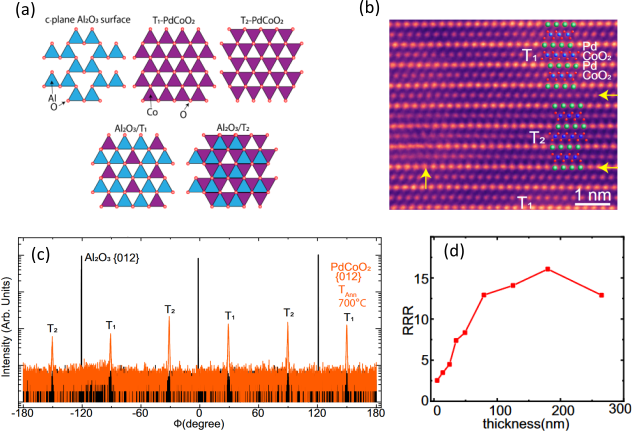}
    \caption{(a) In-plane triangular motifs for Al$_2$O$_3$ and delafossites and the possible twins when films are grown on Al$_2$O$_3$ substrate. (b) Twin domains in the MBE grown films of PdCoO$_2$ seen in STEM. (c) XRD phi scans verify the twin domains in PdCoO$_2$ films. (d) The scaling of RRR with thickness for PdCoO$_2$ films. Reproduced from Brahlek et al \cite{Brahlek2019}. }
    \label{fig:PdCoO2}
\end{figure}

Epitaxial thin films of metallic delafossites have been investigated using PLD \cite{Harada2018,Ok2020}, MBE \cite{Brahlek2019,Sun2019} and chemical methods \cite{Wei2020}. One key MBE growth challenge is due to the noble metals involved: Pd and Pt are difficult to oxidize, and although this can bypassed by the use of proper oxidants the quality of samples are limited by the lower temperature at which growth occurrs. Brahlek et al used oxygen plasma to grow the films at a temperature around 300 $^\circ$C and applied post-growth annealing to help enhance the film quality \cite{Brahlek2019} while Sun et al used distilled ozone and were able to grow at higher temperatures (around 500 $^\circ$C) \cite{Sun2019}. Another challenge is the lack of proper substrates. Unlike the cubic perovskites which have many commercially available substrates, not many lattice-matched substrates are available for the delafossites. The in-plane lattice constant of these materials are in the range of 2.83 - 3.02 $\textrm{\AA}$ which presents a challenge in obtaining a proper lattice-matched substrate. However, high-quality films can still be readily grown on Al$_2$O$_3$ substrate, which has an effective lattice constant of 2.76 $\textrm{\AA}$ and is insulating and chemically inert. The main issue in growing delafossites on Al$_2$O$_3$ substrate is that  there is a large possibility of defects and disorder due to lattice mismatch induced strain. Furthermore, the triangular nature of the $a-b$ plane effectively leads to twin domains which has been verified in MBE-grown films \cite{Brahlek2019,Sun2019} as shown in Figure \ref{fig:PdCoO2}. This type of disorder may enhance the effective mass \cite{barbalas_disorder-enhanced_2022} and is likely the main reason that the epitaxial films do not have a large RRR: the best bulk-like RRR for thick PdCoO$_2$ epitaxial films is about 16 as shown in Figure \ref{fig:PdCoO2}(e), compared to 2000 for bulk single crystals.  More information on thin film metallic delafossites can be found in a recent perspective article \cite{harada_metallic_2022}.

Although some members of the metallic delafossites have been grown successfully with MBE, there are significant challenges in properly stabilizing other members. In the case of PdCrO$_2$, the main challenge is the formation of the Cr$_2$O$_3$ impurity phase, which severely hampers the overall growth of this material. Cr$_2$O$_3$ is structurally similar to Al$_2$O$_3$ and has low oxidation potential, which leads to a very narrow growth window for PdCrO$_2$ films \cite{Ok2020}. Good quality PdCrO$_2$ can only be stabilized by using CuCrO$_2$ as a buffer layer, which can be readily done in MBE \cite{Shin2012,rimal_diffusion-assisted_2022}, and by using proper oxidation and layer control, films with resistivity of around 50 $\mu\Omega cm$ were obtained \cite{song_surface_2024}. PtCoO$_2$ is also challenging, mainly due to the low vapor pressure and high oxidation potential of Pt. The nucleation of stable PtCoO$_2$ phase was found to be a limiting factor, and a buffer layer of PdCoO$_2$ was necessary to obtain single-phase MBE-grown films \cite{song_surface_2024}. New methods have now enabled bypassing the buffer layer and films with better figures of merit have been obtained \cite{li_ptcoo2_2025}.   

For applications, these oxides have been investigated for compatibility with other technologically relevant hexagonal materials. The main application for these films was as a metallic contact for use with Ga$_2$O$_3$ in high temperature Schottky devices \cite{Harada2019a} and as transparent conductors \cite{Harada2018}. MBE-grown epitaxial films of PdCoO$_2$ have shown strong plasmonic enhancements in the terahertz and mid-infrared and are potentially applicable for sensing and nonlinear optical applications \cite{macis_infrared_2022}. The large Rashba splitting and surface magnetism in PdCoO$_2$ have led to proposed applications in spintronics and similar technologies \cite{lee_nonreciprocal_2021}. In other applications, PdCoO$_2$ is used as catalysts in oxygen and hydrogen evolution reactions \cite{Li2019,Liu2022}.

\section{Outlook }
Even though ultra-high conductivity materials with outstanding properties have been realized using MBE, challenges still persist. For some of these materials the resistivity values are much higher than bulk ones, and the residual resistivity is also typically orders of magnitude higher which has inhibited applications in low temperature research. For materials that have been stabilized through MBE, more work is needed to develop solutions that can enable better quality materials with lower defects. Developments are also needed in the growth of high quality epitaxial films for other materials. For example, ReO$_3$ is an interesting ultra-high conductivity material that has yet to be realized using MBE. Some, such as CrO$_2$, are inherently difficult to stabilize in MBE. 

\begin{figure}[ht]
\includegraphics[width=0.9\linewidth]{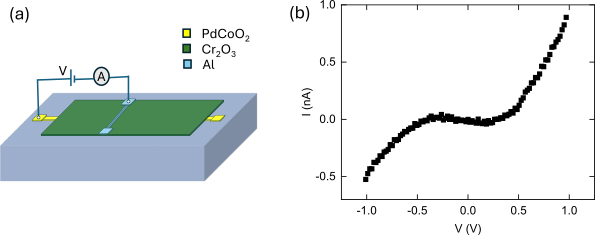}
\caption{(a) Device schematic showing the sandwich structure of PdCoO$_2$-Cr$_2$O$_3$-Al. (b) The current-voltage characteristics of the capacitive structure. }
\label{fig:PdCoO2_device}
\end{figure}

Advancements in semiconductor technology demand innovative materials that can meet the challenges posed by ever-shrinking device sizes. Metallic oxides with minimal defect densities present a compelling alternative to traditional metal interconnects in the semiconductor industry, particularly given the growing difficulty of identifying new, stable metals for this application. Due to their robustness in ambient conditions, these oxides not only represent an excellent choice for interconnect materials but can also function effectively as metal electrodes for back-gating various semiconductor materials.

A significant advantage of metallic oxides over conventional metals lies in their compatibility with single-crystalline dielectrics. These dielectrics can be epitaxially grown on thin films of metallic oxides, enabling high-quality semiconductor materials to be subsequently deposited on top. This bottom-up stacking approach comprising metallic oxides, dielectrics, and semiconductors opens up exciting avenues for research and development, paving the way for applications in advanced semiconductor technologies.

As a perspective of this approach, we present a simple demonstration of the current-voltage characteristics of a PdCoO$_2$-Cr$_2$O$_3$-Al structure. The device schematic is illustrated in Figure \ref{fig:PdCoO2_device}. The fabrication process begins with the epitaxial growth of PdCoO$_2$ on Al$_2$O$_3$ substrate via MBE \cite{Brahlek2019}. The resultant film is then lithographically patterned and dry-etched into strip size measuring 10 $\mu$m. Following this, 100 nm amorphous Cr$_2$O$_3$ is deposited on top of the PdCoO$_2$ by MBE at room temperature. Finally, aluminum is sputtered and patterned into strips having a width of 5 $\mu$m, resulting in capacitive sandwich structure with the contact area of 50 $\mu$m$^2$. The capacitive structure exhibits an impressive resistance of approximately 5 G$\Omega$ at low sourcing voltages ($<$ 0.5 V), underscoring the viable application of metallic oxides as effective electrodes.

There are also opportunities for epitaxial integration of these materials for understanding and applying magnetic and proximity couplings between magnetic metallic oxides such as PdCrO$_2$, and any epitaxially grown materials with compatible lattice structures like hexagonal topological materials. Similar avenues can be explored in materials that exhibit other novel behavior such as altermagnetism, superconductivity or strong correlations. This opens exciting new research pathways that could further advance the field of microelectronic and quantum technologies. For quantum technologies, much research has now focused on applying materials beyond Al for developing superconducting qubits \cite{de_leon_materials_2021} as well as the quest for the mysterious Majorana fermions to construct topological qubits. For the latter approach, topological materials are combined with superconductors, and given the variety of physics in oxide materials and their easy compatibility, understanding the fundamental science and development of topological qubits may be a future avenue worth exploring.

\section{Acknowledgments} 
This work was supported by the U. S. Department of Energy, Office of Science, Basic Energy Sciences, Materials Sciences and Engineering Division (MB, SK) and by the U.S. Department of Energy, Office of Science, National Quantum Information Science Research Centers, Quantum Science Center (DM). Work at the University of Delaware was supported by the U.S. Department of Energy (DOE), Office of Science, Basic Energy Sciences (BES), under Award DE-SC0023478 (BO) and by the National Science Foundation (NSF) under Award DMR-2527684 (TT and RBC). Work at Rutgers is supported by Army Research Office’s MURI W911NF2020166, and the center for Quantum Materials Synthesis (cQMS), funded by the Gordon and Betty Moore Foundation’s EPiQS initiative through Grant GBMF10104.

\bibliography{main}

\end{document}